\documentclass[twocolumn,prd,superscriptaddress,amssymb,nofootinbib,amsmath,nobibnotes,aps,prlgroupedaddress]{revtex4-2}  
\usepackage{graphicx}  
\usepackage{dcolumn}   
\usepackage{bm}        
\usepackage{amssymb}   
\usepackage[]{amsmath,amssymb}
\usepackage{graphics,epsfig}
\usepackage{float}
\usepackage{mathtools}
\usepackage{esint}
\usepackage[utf8]{inputenc}
\usepackage[english]{babel}
\usepackage{mathtools}
\usepackage{xcolor}
\usepackage{float}  
\usepackage{setspace} 
\usepackage{indentfirst}  
\usepackage{cancel}  
\usepackage{soul}
\usepackage{hyperref}  
\hypersetup{
            colorlinks=true,
            linkcolor=black,
            urlcolor=blue,
            } 
\usepackage{tcolorbox}   
\newcommand{\Hconf}{\mathcal{H}}

\usepackage{orcidlink}

\begin{document}

\title{Cosmological Gravitational Waves from Ultralight Vector Dark Matter}

\author{Tomas Ferreira Chase \orcidlink{0009-0001-0286-2136}}
\email{tferreirachase@df.uba.ar}
\affiliation{Universidad de Buenos Aires, Facultad de Ciencias Exactas y Naturales, Departamento de Física. Buenos Aires, Argentina.} 
\affiliation{CONICET - Universidad de Buenos Aires, Instituto de Física de Buenos Aires (IFIBA). Buenos Aires, Argentina}

\author{Diana L\'opez Nacir \orcidlink{0000-0003-4398-1147}}
\email{dnacir@df.uba.ar}
\affiliation{Universidad de Buenos Aires, Facultad de Ciencias Exactas y Naturales, Departamento de Física. Buenos Aires, Argentina.} 
\affiliation{CONICET - Universidad de Buenos Aires, Instituto de Física de Buenos Aires (IFIBA). Buenos Aires, Argentina}

\date{\today}

\begin{abstract}
    We compute the abundance of cosmological gravitational waves produced during the evolution of an ultralight vector (spin-1) dark matter field. 
    A homogeneous background vector field breaks spatial isotropy, requiring a Bianchi I geometry and inducing a mixing between the scalar, vector, and tensor perturbation sectors. 
    We derive the perturbation equations in this background and show that, as a consequence of this mixing, scalar perturbations act as a source of tensor modes, generating a stochastic GW background. 
    The production and cosmological evolution of these gravitational waves are implemented in \texttt{class.VFDM}, a modified version of \texttt{CLASS}, from which we obtain the present-day spectrum.
\end{abstract}

\maketitle

\section{Introduction}

A large number of observations support  the existence of Dark Matter (DM).  According to cosmological observations, such as the  cosmic background radiation  \cite{Planck_2018},  approximately $25\%$ of the energy content of the universe  can be explained by this dark component. However, the fundamental particle nature of DM remains unknown. Many models are being studied to explain the dark sector, covering a broad range of masses for the DM particles.  
Ultralight  DM models have recently gained attention due to their characteristic predictions on small scales \cite{Ferreira_2021}. We focus here on ultralight DM particles with masses in principle in the range  ($10^{-28}$,1) eV. The cosmological predictions depend on how the DM is produced in the early universe. We assume that the production mechanism led to a sufficiently coherent and homogeneous state, such that the DM (in the radiation domination era and afterwards) can be described using the classical field approximation.  Such field could be produced during inflation (see for instance \cite{Marsh_2016, Nakayama:2019rhg,Kaneta:2023lki,Kitajima:2023fun}).
We suppose that the classical field can be decomposed into a background homogeneous field,  plus inhomogeneous perturbations. For a different phenomenology associated to other production mechanisms see for instance \cite{Amin:2022nlh}.
  
One of the questions regarding ultralight models is whether the characteristic predictions will let us differentiate the spin of the dark matter field. 
The assumption that a  spin 0 field can account for the total DM abundance  has been widely studied in the literature, and their  masses have been constrained using cosmological data \cite{Marsh:2015wka, Marsh_2016, Rogers:2020ltq}.   
If the DM abundance is assumed to originate from higher-spin fields, deriving cosmological predictions becomes more challenging. One reason is that the presence of background fields with spins generically breaks the isotropy of the spatial  geometry.  
In the standard $\Lambda$CDM cosmological model, the cosmological perturbations at linear order can be decoupled into a scalar, vector and tensor (SVT) sectors that evolve independently. The decoupling is a consequence of the spatial isotropy of the background. When the DM abundance is given by a background vector field with a fixed spatial direction, the isotropy assumption is broken, and the standard SVT decoupling does not occur. In such ultralight vector field DM models, hereafter referred to as the VFDM models,    the evolution of the background field is determined by two time-scales, given by the mass of the boson and the Hubble parameter. In general, when the Hubble parameter is much greater than the boson mass, the field performs rapid oscillations, which are usually averaged out in cosmological contexts. As has been shown in Refs. \cite{Cembranos:2012kk, Cembranos:2013cba}, the  averaging process washes out the anisotropies in the stress-energy tensor of the ultralight fields, making VFDM models viable dark matter candidates.
However, before the VFDM field starts oscillating, the anisotropies might leave an imprint on the cosmological observables. For example,  an anisotropic imprint remains on the matter power spectrum for the modes that enter the horizon while being relativistic \cite{Chase:2024wsq}. To account for such anisotropies the standard cosmological metric has to be extended to  Bianchi I metrics \cite{Chase:2023puj}. 
In VFDM models, the SVT sectors are coupled through the vector field  and the metric shear $\sigma_{ij}$, which is a tensor that describes the anisotropies in a Bianchi I background. For this reason,  the implementation of these models in Boltzmann-Einstein solvers becomes more involved. The implementation presented in Ref. \cite{Chase:2024wsq} of the VFDM model in \texttt{CLASS}, named \texttt{class.VFDM} \cite{github}, focused  only on  the evolution of the scalar sector of Einstein equations and neglected vector and tensor perturbations. In this paper we extend previous results by studying the three SVT sectors and the impact of accounting for their mixing. 
In particular, we present a numerical implementation in \texttt{class.VFDM} of the tensor modes and the results for the gravitational wave spectrum in the VFDM model. 

The paper is structured as follows. In Sec. \ref{sec:Vector field model} we present the VFDM model and  the linear equations needed to evolve cosmological perturbations including the vector and tensor sectors. In Sec. \ref{sec:SVT_dec_early_times} we solve Einstein equations and the vector field equations of motion for the modes that are outside the horizon ($k\tau\ll 1$). We do this by using   Weinberg construction of the adiabatic mode  \cite{Weinberg_2003, Weinberg:tensor_modes} (see Sec. \ref{sec:WeinbergAD}).  
In Sec. \ref{sec:evolution of tensor perturbations}
we focus on the evolution of tensor perturbations. In Sec. \ref{subsec:num_solutions}, we present the numerical results obtained by an implementation of the equations for the tensor perturbations in \texttt{class.VFDM}. The main outcome is the prediction of the  power spectrum of the tensor modes produced due to the SVT mixing (shown in Fig. \ref{fig:GW_spectrum}).  
In Appendix \ref{appendix:background_equations}, we review the background evolution of the VFDM in Bianchi I. In Appendix \ref{appendix:synch_gauge}, we present the results of the main text in Synchronous gauge, which is needed for the numerical implementation. Finally, in Appendix \ref{appendix:class_implementation} we present the details of the implementation in \texttt{class.VFDM}.

\section{Vector field dark matter model} \label{sec:Vector field model}

We assume that the whole dark matter content of the universe is described by an ultralight vector field which is minimally coupled to the other species.
We model the dynamics of such vector field   with a Proca action, given by
\begin{equation}
    S = - \int dx^4 \sqrt{-g} \left[\frac{1}{4} F^{\mu\nu} F_{\mu\nu} + \frac{m^2}{2} A^{\mu} A_{\mu} \right]\,,
    \label{action}
\end{equation}
where $F_{\mu\nu}$ is the usual field tensor $F_{\mu\nu} = \nabla_{\mu}A_{\nu} - \nabla_{\nu}A_{\mu}$, $m$ is the mass of the field and $g$ is the determinant of the metric.

The vector equation of motion is obtained by varying the action with respect to the vector field, and is given by 
\begin{equation}
    \nabla_{\mu} F^{\mu \nu} + m^2 A^{\nu} = 0 \,.
    \label{eq_motion_vector}
\end{equation}

In this work we consider that the VFDM candidate   can  be described as a  sum of a non-vanishing homogeneous background field  and an inhomogeneous perturbation $A^{\mu}(\tau,\vec{x}) = A^{\mu}(\tau) + \delta A^{\mu}(\tau,\vec{x})$. We assume a linearly polarized background field, $A^{i}(\tau)=A(\tau)\,\hat{A}^i$ with $\hat{A}^i$  a versor pointing in a given direction. In the next subsection we present the model to linear order in cosmological perturbation theory. 

Since the stress tensor of the VFDM is   anisotropic at the background level, for  consistency we consider a Bianchi type I spacetime, with metric given by
\begin{equation}
    ds^2 = a(\tau)^2 \left[ - d\tau^2 + \,\gamma_{ij}(\tau) \, dx^i dx^j \right]\,,
\end{equation}
where $\gamma_{ij} = e^{-2\beta_i(\tau)} \delta_{ij}$ and the functions  $\beta_i$ (with $i=1,2,3$) are constrained by $\sum_i^3 \beta_i = 0$. Following \cite{Pereira_2007}, we define the shear of the metric as $\sigma_{ij} = \frac{1}{2}\gamma_{ij}^{\prime}$, where prime derivatives are with respect to conformal time $\tau$.

For a review of the background evolution see Appendix \ref{appendix:background_equations}. In what follows, we summarize the background behavior at early times. This will be useful for the discussion presented in Sec. \ref{sec:WeinbergAD}.  In order to solve the vector field equations we assume the metric anisotropies (which are generated by the vector field) can be treated perturbatively. We thus neglect the metric shear to solve for the background VFDM.  At early times,  the vector field can be solved as $\vec{A}(\tau) = c \tau \hat{A}$, where   $c$ is a constant determined by the observed cosmological parameters (see Eq. \ref{constant_vector_ini}). This solution is valid until the time of oscillation, defined as the time when $m = H(a_{osc})$.  With the field solution, we can  calculate the vector field fluid variables, given in Eqs. (\ref{cond_ini_background_fluid_1}) and (\ref{cond_ini_background_fluid_2}). 
Then, we can solve for the metric shear using Einstein spatial-traceless equation\footnote{This equation is only sourced by the anisotropic stress of the VFDM,   since it  is the only species that has anisotropies at the background level.}. 
We solve perturbatively for the metric shear, with the vector field anisotropic stress at zero order in this quantity, and acting as a source term on the right-hand side (rhs) of the equation (Eq. \eqref{metric_shear_eq}). In this way, we obtain the solution for the metric shear given by \cite{Chase:2023puj}
\begin{equation}\label{sigmaT}
    \sigma_{ij} = -\frac{c^2}{a^4} \left(\hat{A}_i\hat{A}_j - \frac{\delta_{ij}}{3}\right)\,.
\end{equation}

\subsection{Linear perturbations} \label{sec:Vector field perturbations}

In this section we focus on Einstein equations at linear order in perturbation theory.  We work with the perturbations in Fourier space, using the convention
\begin{equation}
    f(\vec{x}) = \int d^3k\,f(\vec{k})\,e^{i\,\vec{k}\cdot\vec{x}}\,.
\end{equation} 

To work out the perturbations, we use an orthonormal basis given by $\{ \hat{e}_1, \hat{e}_2, \hat{e}_3 \}$, where $\hat{e}_3 = \hat{k}$, $\hat{e}_2 = \hat{k} \times \hat{A}$ and $\hat{e}_1 = \hat{k} \times \hat{e}_2$.  This is a mode-dependent basis where the background vector is always contained in the plane defined by $\hat{k}$ and $\hat{e}_1$. We decompose the metric shear in this basis as (see Eq. (\ref{cond_ini_shear}) for an expression of each component)
\begin{equation}
    \sigma_{ij} = \frac{3}{2} \left( \hat{k}_i \hat{k}_j - \frac{\gamma_{ij}}{3} \right) \sigma_{\parallel} + 2\,\sum_{a=1,2} \sigma_{v_a} \, \hat{k}_{(i}\,\hat{e}_{j)}^a + \sum_{\lambda=+,\times} \sigma_{\lambda} \,\epsilon_{ij}^{\lambda} \,,
\end{equation}
where $\epsilon_{ij}^{+} = \hat{e}_i^1 \,\hat{e}_j^1 - \hat{e}_i^2 \,\hat{e}_j^2$ and $\epsilon_{ij}^{\times} = \hat{e}_i^1\, \hat{e}_j^2 + \hat{e}_i^2 \,\hat{e}_j^1$. Furthermore, we decompose the vector field perturbations in this basis as
\begin{equation}
    \delta \vec{A} = \delta A_L \hat{k} + \delta A_{t_1} \hat{e}_1 + \delta A_{t_2} \hat{e}_2\,.
\end{equation}

We use that the vectors $k_i$ satisfy $\left( k_i \right)^{\prime} = 0$, while (taking into account that $\gamma^{ij}$ is time-dependent) the rest of the relevant versors can be derived as
\begin{subequations}
\begin{align}
    k^{\prime}\,\,\,\, &= - \sigma_{\parallel} \, k\,, \\[6pt]
    (\hat{k}_i)^{\,\prime} &= \sigma_{\parallel} \, \hat{k}_i\,, \\[6pt]
    (\hat{e}^a_i)^{\,\prime} &= - \sum_b \sigma_{lj} e_a^l e_b^j \, e^b_i + 2 \sigma_{ij} e^j_a \,.
\end{align}
\end{subequations}

We work here in Newtonian gauge, and we  present the equations in Synchronous gauge in Appendix \ref{appendix:synch_gauge}. In a Bianchi I background, it is convenient to parameterize the metric linear perturbations as \cite{Pereira_2007}
\begin{subequations}
\begin{align}
    \delta g_{00} &= -2 a^2 \phi\,,\\[3pt]
    \delta g_{0i} &= a^2 B_i \,,\\[3pt]
    \delta g_{ij} &= -2 a^2 \left(\gamma_{ij} + \frac{\sigma_{ij}}{\Hconf}\right) \psi\, + 2 a^2 E_{ij} \,,
\end{align}
\end{subequations}
with ${B_i}^{,i} = 0$ and ${E^i}_i = 0$, and $\Hconf=a H$ is the conformal Hubble rate. The previous definition of the spatial metric, written explicitly in terms of the background shear tensor, makes the tensor perturbations gauge invariant. We parameterize the tensor perturbations in terms of the usual scalar functions $h_+$ and $h_{\times}$, as
\begin{equation}
    E_{ij} = \epsilon_{ij}^{+}\, h_{+} +  \epsilon_{ij}^{\times}\, h_{\times}\,.
    \label{tensor_decomposition}
\end{equation}

The equation of motion for the vector field, Eq. (\ref{eq_motion_vector}), can be split into a constraint equation for the temporal component, given by
\begin{align}
    \delta A_0 = &\frac{- i  k}{m^2 a^2 + k^2} \left[\delta A_L^{\prime}- A_L^{\prime} (\phi + \psi)  + \frac{m^2 a^2}{k} A_{t_1} V_1 \right]\,, \label{Eq_vinculo_deltaA0} 
\end{align}
and an equation of motion for each spatial component. 
By projecting the spatial equation on the basis $\{\hat{k}, \hat{e}_1,\hat{e}_2\}$, and extracting the linear order, we obtain respectively
\begin{subequations} 
\begin{align}
    \delta A_L^{\prime\prime} + m^2 a^2  \delta A_L - ik \delta A_0^{\prime} &= S_L \,,
    \label{eq_mov_delta_A_L}\\[6pt]
    \delta A_{t_1}^{\prime\prime} + (m^2 a^2 + k^2 ) \delta A_{t_1} &= S_{t_1} \,,
    \label{eq_mov_delta_A_T_1}\\[6pt]
    \delta A_{t_2}^{\prime\prime} + (m^2 a^2 + k^2) \delta A_{t_2} &= S_{t_2} \,,
    \label{eq_mov_delta_A_T_2}
\end{align}
\label{eq_mov_delta_A}
\end{subequations}
where the sources on the rhs are given by
\begin{subequations}
\begin{align}
    S_L &= - 2 m^2 a^2 A_{L} \phi + A_L^{\prime} \left(\phi^{\prime} + \psi^{\prime} \right)\,, \\[4pt]
    S_{t_1}  &= - 2 m^2 a^2 A_{t_1} \phi + A_{t_1}^{\prime} (\phi^{\prime} + \psi^{\prime})\\[3pt]
    &\,\quad- ik\,A_L^{\prime} V_1 + 2 A_{t_1}^{\prime} h_+^{\prime}\,, \nonumber\\[4pt]
    S_{t_2} &= 2 A_{t_2}^{\prime} h_{\times}^{\prime} - i k A_{L}^{\prime} V_2 \,.
\end{align}
\end{subequations} 
We thus see that the SVT sectors are coupled  even at zero order in the metric shear. 

\section{Evolution outside the  horizon} \label{sec:SVT_dec_early_times}

In this section we solve Einstein equations and the vector field equations of motion for the modes outside the horizon, defined as $k\tau \ll 1$.
We obtain the adiabatic mode for the cosmological perturbations using  Weinberg's construction. We show that it is a solution of Einstein equations  in the $k\to0$ limit at leading order in the anisotropies, due to a cancellation between contributions involving the vector field and the shear tensor. As it will be enough, we retain corrections up to linear order in the metric shear in the Einstein tensor, and up to zeroth order in the VFDM stress–energy tensor.

\subsection{Adiabatic mode for the vector field from   Weinberg's construction }  \label{sec:WeinbergAD}

In Refs. \cite{Weinberg_2003, Weinberg:tensor_modes} the author uses a residual symmetry of Einstein equations at $k=0$ to find approximate solutions for the modes that are outside the horizon. In the $\rm{\Lambda CDM}$ scenario, the linearized Einstein equations in Newtonian gauge are invariant under a redefinition of the time coordinates and a rescaling of spatial coordinates for $k=0$, namely
\begin{equation}
    t \to t + \epsilon(t)\,, \qquad  x^i \to ({\delta^i}_j - \frac{1}{2}{w^i}_j) x^j\,,
    \label{gauge_transformtion}
\end{equation}
where $\epsilon(t)$ is an arbitrary function of time and where $w_{ij} = w_{ji}$ is an arbitrary constant matrix.
As was shown in the previous references, this symmetry can be used to construct solutions of Einstein equations in the $k\to0$ limit. This construction works for effective fluids with vanishing anisotropic stress as $k\to0$.

In Ref. \cite{Chase:2023puj} it was shown that, even though the VFDM is a subdominant component   in the early universe, it is non-trivial to assess whether the anisotropies generated by the vector field could significantly break the invariance under (\ref{gauge_transformtion}) or not. Neglecting vector and tensor perturbations, the key results of such reference are as follows.  On one hand, some of the Einstein equations (the temporal equation, the  spatial-temporal  and the spatial equation proportional to $\delta_{ij}$)   reduce to the usual form with a FRW background outside the horizon, provided the shear and the contribution from the VFDM in radiation-dominated era can be neglected.  On the other hand, the spatial equation that is not proportional to $\delta_{ij}$ has non-negligible contributions at $k=0$. These contributions involve perturbations to the Bianchi background and the vector field anisotropic stress. However, it turns out that such contributions are canceled at leading order in $\sigma_{ij}$ for adiabatic modes. Therefore, the solutions obtained using Weinberg's construction are a good approximation even in the presence of a VFDM. In the next sub-section we show that this is still the case  when considering tensor modes.

In order to derive the adiabatic mode for the vector field, we follow Ref. \cite{Weinberg:tensor_modes}. 
We hence use that the Weinberg's construction holds at leading order in the anisotropies, and apply the transformation  in Eq. (\ref{gauge_transformtion})   to obtain the adiabatic mode.   
By performing such transformation  on the metric, on the vector field, and on  the fluid variables of  each species, we find the corresponding solutions for the system at $k=0$, obtaining 
\begin{subequations}
\begin{align}
    \delta A_i &=   - \epsilon   \dot{A_i} + \frac{1}{2} w_{ij} A^j \,,\label{WAM_A}\\[3pt]
    \phi &= - \dot{\epsilon}\,, \label{WAM_phi}\\[6pt]
    \psi &= H \epsilon - \frac{w}{6}\,,\label{WAM_psi}\\[6pt]
    E_{ij} &= \frac{w_{ij}}{2} - \frac{w}{6} \gamma_{ij} - \frac{\sigma_{ij}}{\Hconf} \frac{w}{6}\,, \label{WAM_tensor} \\[6pt]
    \delta g_{0i} &= 0\,,\\[6pt]
    \delta \rho_s &= -\dot{\rho}_s \epsilon\,,\label{adiabatic_conds_ini}\\[6pt]
    \delta P_s &= -\dot{P}_s \epsilon \,,
\end{align}
\label{fluid_adiabatic_ini_cond}
\end{subequations}where the subscript $s$ runs for each species, $w = {w^k}_k$ and  a dot represents a derivative with respect to cosmic time. 
It can be shown that the   previous expressions are a solution of the equations in the limit $k\to 0$, provided the anisotropic stress scales as $k^2$ in such limit. 
Furthermore, if the anisotropic stress scales as $a^2$ in such limit, it can be shown that $\dot{\epsilon}=const$ either in radiation or matter era, so $\psi$ and $\phi$ are also constant. This is the case for the radiation components in radiation dominated era. 

Now we focus on the solution for the tensor perturbations.
We can project \eqref{WAM_tensor} using $\epsilon^+_{ij} = \hat{e}_1^i\hat{e}_1^j - \hat{e}_2^i\hat{e}_2^j$, which satisfies
\begin{equation}
    \epsilon_+^{ij} h_{ij} = 2 h_+\,,\quad \epsilon_+^{ij} w_{ij} = 2w_+\,, \quad \epsilon_+^{ij} \gamma_{ij} = 0\,. 
\end{equation}
In this way, we extract the solution for $h_+$, and in an analogous way for $h_{\times}$, as
\begin{align}
    h_a &= \frac{w_a}{2} + \sigma_a \left[ \frac{\psi}{\Hconf} - \frac{\psi^{\prime} + \Hconf\phi}{\Hconf^{\prime} - \Hconf^2}\right]\,,
    \label{WAM_h_+}
\end{align}
where $a = +, \times$.
In the radiation era and for the modes outside the horizon, the tensor perturbations reduce to 
\begin{equation}
    h_a = \frac{w_a}{2} + \frac{\sigma_a}{\Hconf}(\psi + \frac{\phi}{2})\,.
    \label{WAM_h_+_RAD}
\end{equation}
Before the field starts oscillating, the previous expression is approximately constant. After the field starts oscillating, the metric shear starts decaying, so $h_a$ tends to a constant ($w_a$) fixed by the initial conditions. 

Now we calculate the initial condition for the vector field. Writing \eqref{WAM_A} in terms of the metric perturbations using \eqref{WAM_phi}, \eqref{WAM_psi} and \eqref{WAM_h_+}, and using the fact that in the radiation domination era the background field satisfies  $\dot{A_i} \simeq {H}A_i$, the vector field can be expressed initially as
\begin{align}
    \delta A_i &= - A_i \psi + \frac{1}{2} A^k E_{ki}\nonumber\\[4pt]
    &= - A_L \psi \,\hat{k} - A_{t_1} (\psi - h_+)\, \hat{e}_1 + A_{t_1} h_{\times}\, \hat{e}_2\,.
    \label{vector_cond_ini}
\end{align}
We notice that Eq. \eqref{WAM_A} satisfies the VFDM equations of motion for $k\to0$ and to zero order in the metric shear, not only in radiation dominated era but also afterwards.

\subsection{SVT decomposition and initial conditions}

We now present the scalar, vector, and tensor sectors of Einstein equations in Newtonian gauge. 
We focus on super-horizon scales, relevant for the initial conditions of the perturbations.
We show explicitly that the adiabatic mode constructed in the previous section solves the corresponding equations in the $k\to 0$ limit.

\subsubsection{Scalar sector}

For a perturbed energy momentum tensor ${\delta T^{\mu}}_{\nu}$,    scalar effective fluid variables can be defined  as:
\begin{subequations}
\begin{align}
    \delta\rho &= - \delta {T^0}_0 \,,\\[6pt]
    \delta P &= \frac{1}{3}\delta {T^i}_i\,, \\[6pt]
    (\rho + P) \theta &= i k^i \delta {T^0}_i \,,\\[6pt]
    (\rho + P) \delta\Sigma_{\parallel} &= - \big(\hat{k}_i\hat{k}^j - \frac{1}{3} {\gamma^j}_i\big) \delta {\Sigma^i}_j\,,
\end{align}
\label{fluid_variables_definitions}
\end{subequations} 
where $\delta{\Sigma^i}_j = \delta{T^i}_j - \delta{T^k}_k {\delta^i}_j/3$ is the anisotropic stress. 
For the VFDM, the fluid variables in Newtonian gauge are given by
\begin{subequations}
\begin{align}
    \delta \rho_A &= \frac{1}{a^4} \big[A_L^{\prime}\left(\delta A^{\prime}_L - i\,k\,\delta A_0\right)  +  A_{t_1}^{\prime} \delta A_{t_1}^{\prime}  \label{delta_rho_A}\\[3pt]
    &+ m^2 a^2 \left(A_L \,\delta A_L+ A_{t_1} \,\delta A_{t_1}\right) \big] + 2\rho_A \, \psi  \nonumber\\[3pt]
    &- 2 \sin(\gamma_k)^2\rho_A h_+ - (\rho_A + 3 P_A) \phi\,. \nonumber
\end{align}
\begin{align}
    \delta P_A &= \frac{1}{3 a^4} \big[A^{\prime}_L \left(\delta A^{\prime}_L - i\,k\,\delta A_0\right) + A_{t_1}^{\prime} \delta A_{t_1}^{\prime} \label{delta_P_A}\\[3pt]
    &- m^2 a^2 \left(A_L \,\delta A_L+ A_{t_1} \,\delta A_{t_1}\right)  \big] + 2P_A \psi \nonumber\\[3pt]
    &- 6 \sin(\gamma_k)^2 P_A h_+ - \frac{1}{3}(\rho_A + 3 P_A) \phi  \,, \nonumber
\end{align}
\begin{align} 
    (\rho_A + P_A) \theta_A &=  \frac{k^2}{a^4} A^{\prime}_{t_1}  \delta A_{t_1} -i \frac{m^2 \, k}{a^2} \, A_L\, \delta A_0 \\[3pt]
    & + i \frac{m^2 k}{a^2} A_L A_{t_1} V_1\,, \nonumber
\end{align}
\begin{align}
   (\rho_A &+ P_A) \delta\Sigma_{A\parallel}  = \frac{2}{3 a^4} \big[2A^{\prime}_L(\delta A^{\prime}_L - i\,k\,\delta A_0) \label{shear_A}\\[3pt]
   &- A_{t_1}^{\prime} \delta A_{t_1}^{\prime} + m^2 a^2 \big(A_{t_1} \,\delta A_{t_1} - 2 A_L \,\delta A_L\big)  \big]  \nonumber\\[3pt]
    &+ 2 \,\Sigma_{\parallel}\,\psi + \frac{2}{3 a^4}\left[A_{t_1}^{\prime\,2} - 2 A_L^{\prime\,2}\right] \phi - 4 \sin(\gamma_k)^2 P_A h_+ \nonumber\,, 
\end{align}
\label{fluid_variables_Newtonian}
\end{subequations}
where  $\cos(\gamma_k) = \hat{k}\cdot\hat{A}$. We note that the tensor perturbations decouple for a purely longitudinal Fourier mode ($\gamma_k = 0$).

Now we present the scalar sector of the Einstein tensor, which in Newtonian gauge  can be written as 
\begin{subequations} 
\begin{equation}
    \delta{G^0}_0 = k^2 \psi + 3 \Hconf (\psi^{\prime} + \Hconf \phi) + \frac{\sigma_+}{a^2} h_+^{\prime} - \frac{i k}{a^2} \sigma_{v_1} V_1  \,, \label{Einstein00Newtonian}
\end{equation}
\begin{equation}    
    \hat{k}^i \delta{G^0}_i = k^2\, (\psi^{\prime} + \Hconf\phi) - \frac{k^2}{a^2} \sigma_+ h_+^{\prime} \,,\label{Einstein0iNewtonian}
\end{equation}
\begin{align}
    \delta{G^k}_k &= \psi^{\prime \prime} + (\Hconf^2 + 2 \Hconf^{\prime}) \phi + \Hconf (\phi^{\prime} + 2 \psi^{\prime}) \label{EinsteinSSNewtonian} \\[3pt]
    &+ \frac{k^2}{3} (\psi - \phi) - \frac{\sigma_+}{a^2}\, h_+^{\prime} + i\frac{k}{a^2} \, \sigma_{v_1} V_1 \,,
   \nonumber
\end{align}
\begin{align}
    &\delta G_{\parallel} = k^2 (\psi - \phi) + 3( \sigma_{\parallel}^{\prime} + 2 \Hconf \sigma_{\parallel}) \phi \label{EinsteinSSTTNewtonian}\\[3pt]
    &+\frac{3}{2}\left[ \sigma_{\parallel} \left( 2\frac{\Hconf^{\prime}}{\Hconf} \left(\frac{\Hconf^{\prime}}{\Hconf^2} - 1\right) - \frac{\Hconf^{\prime\prime}}{\Hconf^2} - \frac{k^2}{3} \right) \right.\nonumber\\[3pt]
    &+ \left. \frac{\sigma_{\parallel}^{\prime \prime}}{\Hconf} + 2 \sigma_{\parallel}^{\prime} \left( 1 - \frac{\Hconf^{\prime}}{\Hconf^2} \right) \right] \psi \nonumber + \frac{k}{3 a^2} \sigma_{v_1} V_1\,, \nonumber
\end{align}\label{einstein_eqs_newt}
\end{subequations}
where we kept the expressions to linear order in the background metric shear.

When the SVT decoupling holds, one can calculate the initial conditions for the metric perturbation  by solving for the metric along with the radiation perturbations, since they dominate at early times. In this way, we obtain \cite{Ma:1995ey} 
\begin{subequations}
\begin{align}
    \phi &= \frac{20 \,C}{15+ 4 R_{\nu}}\,,  \\[3pt]
    \psi &= (1+\frac{2}{5} R_{\nu})\phi \\[3pt]
    \delta_{\gamma} &= -2 \phi \,,\\[3pt]
    \theta_{\gamma}  &= \frac{1}{2} k^2 \tau \,\phi \,,
\end{align}
\end{subequations} 
where $C$ is a constant and $R_{\nu} = \frac{\rho_{\nu}}{\rho_{\gamma}}$. The previous expressions are equivalent to the adiabatic solution of Eq. (\ref{fluid_adiabatic_ini_cond}) under the assumption of a vanishing anisotropic stress at $k\to0$.
For the adiabatic mode  the tensor perturbations are constant at early times, and the vector metric perturbations are zero. Therefore, we can neglect the mixing in   Eqs. (\ref{Einstein00Newtonian}), (\ref{Einstein0iNewtonian}) and (\ref{EinsteinSSNewtonian}). In Eq. (\ref{EinsteinSSTTNewtonian}), however, the terms containing the shear tensor have to be kept, since they dominate in the $k\to0$ limit. As we show next, the anisotropic stress of the VFDM does not vanish in the $k\to0$ limit. Instead, it cancels the dominant Bianchi contribution from Eq. \eqref{EinsteinSSTTNewtonian}, which also remains in the $k\to0$ limit. 

In what follows, we calculate the initial conditions for the fluid variables. As can be seen in Eqs. (\ref{fluid_variables_Newtonian}), the scalar and tensor sectors are mixed in the vector fluid variables at zero order in the metric shear strength ($|\sigma_{ij}|$). In contrast with the Einstein tensor, the mixing is non-negligible for $k \ll \Hconf$. However, by imposing the adiabatic mode (Eqs. \eqref{fluid_adiabatic_ini_cond} and \eqref{vector_cond_ini}), the contribution from the tensor perturbations  is  canceled by the vector field terms, thus decoupling the scalar fluid variables from the tensor metric perturbations. We can explicitly check this cancellation, for example, on the vector energy density. The rhs of Einstein temporal equation is given by Eq. (\ref{delta_rho_A}).
By inserting the vector solution of Eq. (\ref{vector_cond_ini}), and working in the regime outside the horizon in the radiation era, we obtain
\begin{align}
    \delta \rho_A \simeq \,&\frac{1}{a^4} (A^{\prime}_L\delta A^{\prime}_L  +  A_{t_1}^{\prime}  \delta A_{t_1}^{\prime}) - \frac{A^{\prime^2}_{t_1}}{a^4} h_+ \nonumber\\[4pt]
    &+ 2\rho_A \, \psi - (\rho_A + 3 P_A) \phi\nonumber\,\\[6pt]
    \simeq &-2 \rho_A \phi\,, \label{cond_ini_delta_A_newt}
\end{align}
where we used $A^{\prime}_i \sim \Hconf A_i \gg m A_i$.
Using that before the field starts oscillating $\rho_A\sim a^{-4}$ (see Eq. (\ref{cond_ini_background_fluid_1})), we can check that the previous expression is the adiabatic initial condition for the vector field, as defined in Eq. (\ref{adiabatic_conds_ini}).
We calculate the initial conditions for the rest of the fluid variables by inserting Eq. (\ref{vector_cond_ini}) into the corresponding expressions in Eq. (\ref{fluid_variables_Newtonian}), obtaining
\begin{align}
    \delta P_A &= -\frac{2}{3} \rho_A \phi \,,\\[3pt]
    (\rho_A+P_A)\delta\Sigma_{\parallel, A} &= \frac{2}{3}\frac{c_{t_1}^2-2c_L^2}{a^4}\,\phi\,,\label{vecshearNewt}\\[3pt]
    (\rho_A+P_A)\theta_A &= \frac{c_L^2 (\phi + 2 \psi) - c_{t_1}^2\psi}{a^4}\, k^2\tau\,.
\end{align}
In the calculation of these  expressions, as for Eq. (\ref{cond_ini_delta_A_newt}), there are   cancellations between the contributions from the tensor perturbation and the vector field. Moreover, we can see that the initial condition for $\delta\Sigma_{\parallel, A}$ does not vanish in the $k\to 0$ limit. As mentioned above, this contribution is   canceled by the Bianchi I correction in (\ref{EinsteinSSTTNewtonian}). So, the constructed adiabatic modes are a consistent approximate  solution in the limit $k\to 0$.

\subsubsection{Vector sector}

The Einstein vector equations can be obtained by projecting $\delta {G^0}_i$ in the directions $\hat{e}_1$ and $\hat{e}_2$, thus obtaining respectively
\begin{align}
    \left[\frac{m^2}{m_{P}^2} A_{t_1}^2 + \frac{k^2}{2}\right] V_1 &= \frac{m^2 A_{t_1}}{m_{P}^2} \delta A_0 + S_V \,,\label{Einstein_vec_V1}\\[5pt]
    \frac{k}{2} V_2  &= - 2 i \sigma_{v_1} h_{\times} - \frac{i}{a^2 m_P^2} A_L^{\prime}\delta A_{t_2}\,,
    \label{Einstein_vec_V2}
\end{align}
where $S_V$ is
\begin{align}
    S_V = &-i k \frac{1}{\Hconf^2}\left[(3\Hconf^2-\Hconf^{\prime})\sigma_{v_1} + \Hconf\sigma_{v_1}^{\prime} \right]\psi \\[3pt]
    &- i k \sigma_{v_1} \phi - i k \frac{\sigma_{v_1}}{\Hconf}\psi^{\prime} \,.\nonumber
\end{align}

The Weinberg construction  of the adiabatic mode excludes  vector perturbations. We initially set this kind of perturbation to zero. To see that the solution with vanishing vector perturbations is consistent in the presence of background anisotropies, we check that a vanishing vector metric perturbation is a solution of the Einstein vector equations in the $k\to0$ limit. As can be seen by inserting (\ref{fluid_adiabatic_ini_cond}) and (\ref{vector_cond_ini}) in Eq. (\ref{Einstein_vec_V1}), and using Eq. (\ref{Eq_vinculo_deltaA0}), the zero order in $k$   vanishes. Then, the equation is satisfied outside the horizon at leading order in the anisotropies. In Eq. (\ref{Einstein_vec_V2}), a similar cancellation occurs between $h_{\times}$ and the vector field on the rhs. 

\subsubsection{Tensor sector} \label{sec:einstein_tensor_early}

The tensor sector is obtained by projecting Einstein equations with
\begin{equation}
    {\Lambda^{il}}_{jm} = P^{il} P_{jm} - \frac{1}{2} P^{i}_j P^{l}_m\,,
\end{equation}
where $P_{ij} = (\gamma_{ij} - \hat{k}_i\hat{k}_j)$.
The equations of motion for the two polarizations of the tensor perturbations are then calculated by contracting the resulting equations in the directions $\hat{e}^1_i\hat{e}^1_j$ and $\hat{e}^1_i\hat{e}^2_j$. By doing this on both sides of Einstein equation, we obtain for each projection respectively
\begin{subequations}
\begin{align}
    h_{+}^{\prime\prime} &+ 2 \Hconf h_+^{\prime} + \left[ k^2 - \frac{6 a^2}{m_P^2}\sin(\gamma_k) P_{A}  \right] h_+ =  H_+ + \frac{1}{m_P^2} A_+\,, \label{einstein_h_plus}
\end{align}
\begin{align}
    h_{\times}^{\prime\prime} + 2 \Hconf h_{\times}^{\prime} &+ \Big[ k^2 - 2 \sigma_+^{\prime} - 4 \Hconf \sigma_+ - \frac{12 a^2}{m_P^2} \sin(\gamma_k) P_{A} \Big] h_{\times}\nonumber\\
    & = i k \sigma_{v_1} V_1 + \frac{1}{a^2m_P^2} A_{\times} \,,\label{einstein_h_cross}
\end{align} \label{einstein_eqs_tensor_newt}
\end{subequations}
where the sources $H_+$, $A_+$ and $A_{\times}$ are given by
\begin{subequations}
\begin{align}
    A_+ &= \frac{1}{a^2} \left[ m^2 a^2 A_{t_1}  \delta A_{t_1} - A_{t_1}^{\prime} \delta A_{t_1}^{\prime}  \right] + \frac{1}{a^2} {A_{t_1}^{\prime}}^2 \phi\label{einstein_tensor_source_A}\\[5pt]
    &\,\quad- \frac{1}{a^2} ({A_{t_1}^{\prime}}^2 - m^2 a^2 A_{t_1}^2) \psi \nonumber\,, \\[7pt]
    H_+ &= 2(\sigma_+^{\prime} + 2 \Hconf \sigma_+) \phi + \left[\Hconf^{-1}(\sigma_+^{\prime\prime}-(2\Hconf^{\prime}-k^2)\sigma_+)\right. \nonumber\\[5pt]
    &\,\left.+ 2 \sigma_{+}^{\prime} + 2 \Hconf^{-3}{\Hconf^{\prime}}^2 \sigma_+ - \Hconf^{-2}(\Hconf^{\prime\prime} \sigma_+ + 2 \Hconf^{\prime}  \sigma^{\prime}_+)\right]\psi \nonumber \\[5pt]
    &\,-i k \sigma_{v_1} V_1 - \left[ \sigma_+ \left(\frac{2\Hconf^{\prime}}{\Hconf^2} - 5\right)- \frac{2\sigma_+^{\prime}}{\Hconf} \right] \psi^{\prime}\,, \label{einstein_tensor_source_H}\\[7pt]
    A_{\times} &= \frac{1}{a^2}\left[m^2 a^2 A_{t_1}\delta A_{t_2} - A_{t_1}^{\prime} \delta A_{t_2}^{\prime} \right]\,.\label{einstein_tensor_source_cross_A_synch}
\end{align}\label{einstein_tensor_source_synch}
\end{subequations}
We can see that the vector field pressure gives  a mass term for both polarizations of the tensor perturbations. 
By inserting the solution of the perturbations outside the horizon \eqref{fluid_adiabatic_ini_cond} in Eq. (\ref{einstein_eqs_tensor_newt}), two cancellations take place for $a < a_{\rm osc}$. 
Firstly, the mass term for $h_+$ is canceled by the first term in brackets on Eq. (\ref{einstein_tensor_source_A}). 
Secondly, the source from $H_+$, which comes from the mixing of the SVT sectors through the metric shear, is   canceled by the remaining terms in the source $A_+$, which is given by the vector field. 
An analogous result holds for the "$\times$" polarization. 
In this way, we obtain
\begin{align}
    h_{+}^{\prime\prime} + 2 \Hconf h_{+}^{\prime} + k^2 h_+ &= k^2 \frac{\sigma_+}{\Hconf} \psi + i k \sigma_{v_1} V_1\,, \\[5pt]
    h_{\times}^{\prime\prime} + 2 \Hconf h_{\times}^{\prime} + k^2 h_{\times} &=  i k \sigma_{v_1} V_2\,, 
    \label{einstein_tensor_ini}
\end{align}
In particular, for $k=0$ and before the field starts oscillating, we obtain the usual $\Lambda$CDM behavior for the tensor perturbations.

After the field starts oscillating, the cancellations described above are no longer valid. Equation \eqref{fluid_adiabatic_ini_cond} still holds, but the tensor perturbations are no longer constant. Instead, they perform decaying oscillations due to the evolution of $\sigma_{+,\times}$.
If the mode is outside the horizon after the oscillations of the vector field's background are averaged out, the tensor perturbations tend to a constant given by $h_{+,\times} = w_{+,\times}/2$. Notice that only modes that enter the horizon  after the metric shear has significantly decayed  would behave as if there were a primordial gravitational wave with $h_{+,\times} \simeq w_{+,\times}/2$.  Modes that enter the horizon earlier would be affected by the shear term.

\section{Evolution of the tensor perturbations}\label{sec:evolution of tensor perturbations}

In this section we study the evolution of the tensor perturbations and the abundance of gravitational waves produced.
For simplicity, we ignore the photon and neutrino sources. The impact of the radiation over the tensor perturbations is well-known, leading to a damping on the perturbation amplitude \cite{Weinberg:tensor_modes}. Since the vector field does not interact directly with the radiation, we drop such contributions to study the impact of the vector field on the gravitational waves alone.  

Our goal is to predict the gravitational wave abundance at present time, defined as \cite{Caprini:2006jb, Fenu:2009qf, Miron-Granese:2020hyq}
\begin{equation}
    \Omega_{\rm GW}(\vec{k}) = \frac{1}{\rho_c}\frac{d^2 \rho_{GW}}{d\Omega\, d\log(k)} = \frac{k^3}{48 \pi^3 H_0^2}\langle h_{ij}^{\prime \,2}\rangle'\,, 
    \label{GW_abundance_def}
\end{equation}
where $h_{ij}$ indicates the $ij$ component of the gravitational waves, and $\langle\dots\rangle'$ indicates the ensemble average of the field with the $(2\pi)^3  \delta(\vec{k}-\vec{k}^{\prime})$ removed.
The quantity $\langle h_{ij}^{\prime \,2}\rangle$ is calculated from the transfer function of the gravitational waves weighted by primordial power spectrum.

In cosmological models where the SVT sectors are mixed, there could be a non-vanishing primordial  cross-correlation between scalar and tensor modes. We therefore parameterize the '$+$' polarization of the gravitational waves as (for the '$\times$' polarization we use an analogous parameterization)
\begin{equation}
    P_+^{ij}h_{ij}(\vec{k},\tau) = \tilde{h}_{+}(\vec{k})\, \mathcal{T}_{++}(\vec{k}, \tau) + \mathcal{R}(\vec{k})\, \mathcal{T}_{s+}(\vec{k}, \tau) \,.
\end{equation}
In here,  $\mathcal{R}(\vec{k})$ is the curvature perturbation which defines the scalar power spectrum $\mathcal{P}_{s}(\vec{k})$, and $ \mathcal{T}_{s +}(\vec{k}, \tau)$ is the transfer function of the gravitational waves with $+$ polarization produced by  mixing with the scalar sector\footnote{In this parametrization of the tensor perturbations, the normalization of  $\mathcal{T}_{s +}(\vec{k}, \tau)$    is a free parameter that characterizes the primordial  scalar-tensor mixing, which  could  eventually be predicted by a specific inflationary model. Notice $\mathcal{T}_{s +}(\vec{k}, \tau)$ corresponds to the tensor perturbation we called $h_{+}$ in previous sections.}. $\tilde{h}_+(\vec{k})$ defines the  primordial tensor power spectrum $\mathcal{P}_{+}(\vec{k})$, and $ \mathcal{T}_{++}(\vec{k}, \tau)$ is the transfer function for the free evolution (that is, the homogeneous solution of \eqref{einstein_h_plus}). 
We can then compute the power spectrum of the derivative of the amplitude as
\begin{align}
    &\langle h_{+}^{\prime}(\tau, \vec{k})h_{+}^{*\prime}(\tau, \vec{k}^{\prime})\rangle = (2\pi)^3 \delta(\vec{k}-\vec{k}^{\prime})\Big[\mathcal{P}_{+}(\vec{k}) \,|\mathcal{T}^{\prime}_{++}(\vec{k}, \tau)|^2   \nonumber\\[3pt]
    &+ \mathcal{P}_{s}(\vec{k}) \,|\mathcal{T}_{s+}^{\prime}(\vec{k}, \tau)|^2    + \mathcal{P}_{+s}(\vec{k}) \mathcal{T}_{++}^{\prime}(\vec{k}, \tau)\mathcal{T}^{*\prime}_{s+}(\vec{k}, \tau) \nonumber\\[3pt]&+\mathcal{P}^*_{+s}(\vec{k}) \mathcal{T}_{++}^{*\prime}(\vec{k}, \tau)\mathcal{T}^{\prime}_{s+}(\vec{k}, \tau) \Big]\,,
\end{align}
where $\mathcal{P}_{+s}(\vec{k})$ is the power spectrum for the scalar-tensor cross correlation, defined as
\begin{equation}
    \langle \tilde{h}_+(\vec{k})\mathcal{R}^*(\vec{k}^{\prime})\rangle = (2\pi)^3 \delta(\vec{k}-\vec{k}^{\prime})\mathcal{P}_{+s}(\vec{k})\,.
\end{equation} 
To model the primordial power spectra we use
\begin{equation}
    \mathcal{P}_{x}(\vec{k})=(2\pi)^3 k^{-3} \mathcal{A}_x (k/k_p)^{n_x-1}\,,
\end{equation}
where $x$ stands for $s$, $+$, $\times$ or $+s$. Here $\mathcal{A}_x$ is the  amplitude, $n_x$ is the tilt and $k_p$ is a pivot scale.  In standard $\Lambda$CDM cosmology, the scalar and tensor power spectra  are uncorrelated (i.e., $A_{+s}=A_{\times s}=0$), and the  amplitude   is parameterized in terms of the scalar-to-tensor ratio $r$ as  $\mathcal{A}_h=r \mathcal{A}_s$, where $\mathcal{A}_h=4\mathcal{A}_+=4\mathcal{A}_\times$, and the tensor spectral index, denoted as $n_T$, corresponds to setting $n_\times=n_+=1+n_T$. 

In what follows we use the fiducial values for $ n_s$ and $\mathcal{A}_s$   \cite{Planck_2018}, as well as  $\Omega_{\rm cdm}=0$ and $\Omega_A \sim 0.269$. 
For simplicity, we present numerical solutions for the gravitational wave power spectra assuming that there is no primordial power spectrum, meaning that  $\mathcal{A}_+=\mathcal{A}_{\times}=\mathcal{A}_{+s}=\mathcal{A}_{\times s}=0$. 

The presence of the VFDM induces a mixing between the scalar and tensor sectors, so we expect the gravitational waves to produce a backreaction on the scalar perturbations. As shown in Sec. \ref{sec:WeinbergAD}, the super-horizon tensor perturbation $h_+$ is of order $|\sigma_{ij}|$ times a scalar metric perturbation. Therefore, after inserting $h_+$ in the fluid variables given in Eq. (\ref{fluid_variables_Newtonian}), we see the backreaction of such perturbation on the scalar sector is of the same order as the terms linear in $|\sigma_{ij}|$, which we neglected to obtain Eq. (\ref{fluid_variables_Newtonian}) from the VFDM energy-momentum tensor. Therefore, a consistent analysis of the backreaction of tensor perturbations on the scalar sector requires extending all the equations and the code to next-to-leading order in  $|\sigma_{ij}|$. We leave such analysis  for future work, together with its implications for observables such as the CMB multipoles.

\subsection{Numerical solutions}\label{subsec:num_solutions}

In this section we present the results from a numerical implementation of the tensor perturbations in the vector field dark matter model. 
The implementation is based on \texttt{class.VFDM} \cite{github}, our modified version of CLASS \cite{blas2011cosmic}, which solves the scalar sector following Ref. \cite{Chase:2024wsq}. We extend it here to evolve the tensor perturbations coupled to the scalar sector. 
The code is available at \cite{github}. The implementation is performed in Synchronous gauge (see Appendix \ref{appendix:synch_gauge} for the equations). Recall that tensor perturbations are gauge invariant. The details of the implementation are presented in Appendix \ref{appendix:class_implementation}.

Previous numerical solutions for the tensor perturbations with VFDM can be found in Ref. \cite{Miravet:2020kuj}. However, in such reference the authors did not consider adiabatic initial conditions for the vector field, and solved for the vector field dynamics using a WKB approximation in a FRW background, uncoupled from the metric perturbations. In here, we consider the adiabatic initial conditions derived above, and solve for the tensor perturbations coupled to the rest of the species.

\begin{figure*}[th]
    \centering
    \includegraphics[width=0.49\textwidth]{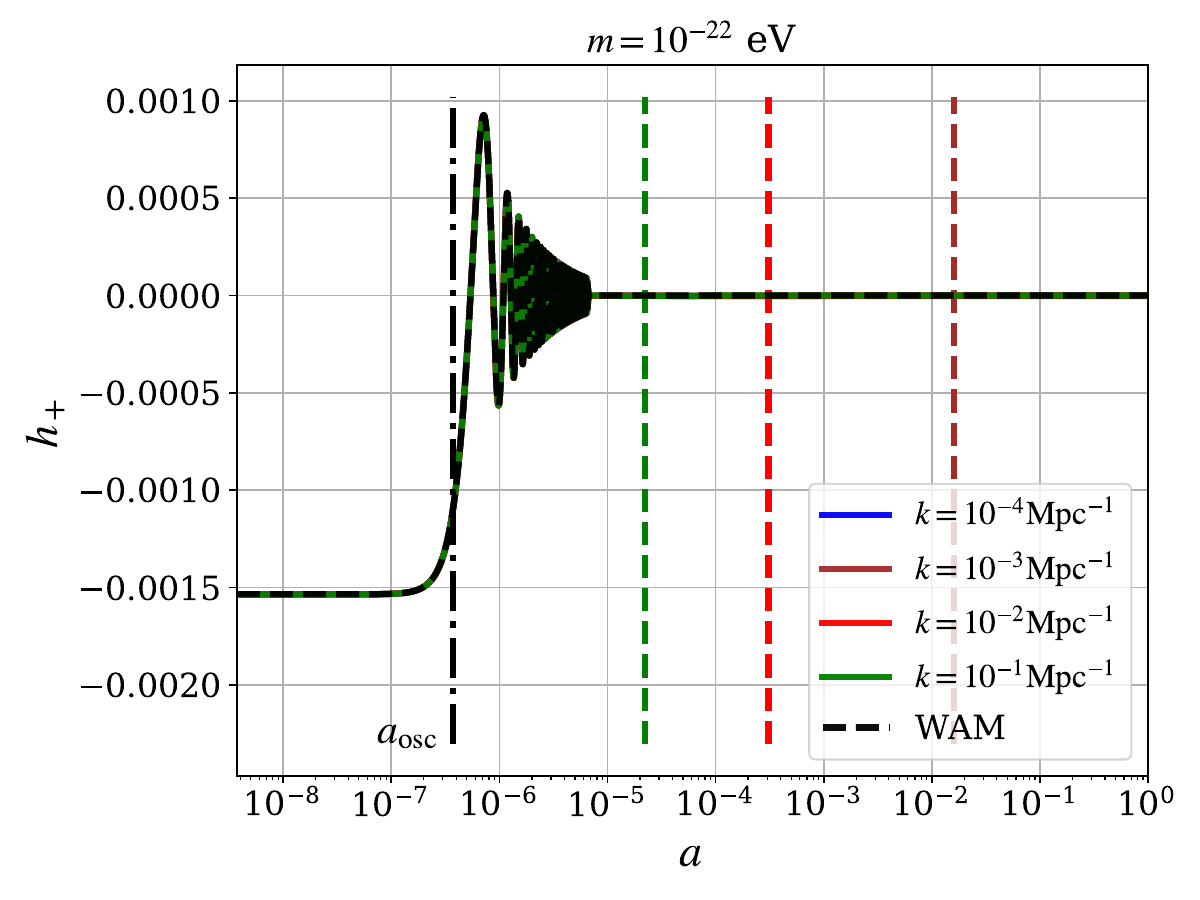}
    \includegraphics[width=0.49\textwidth]{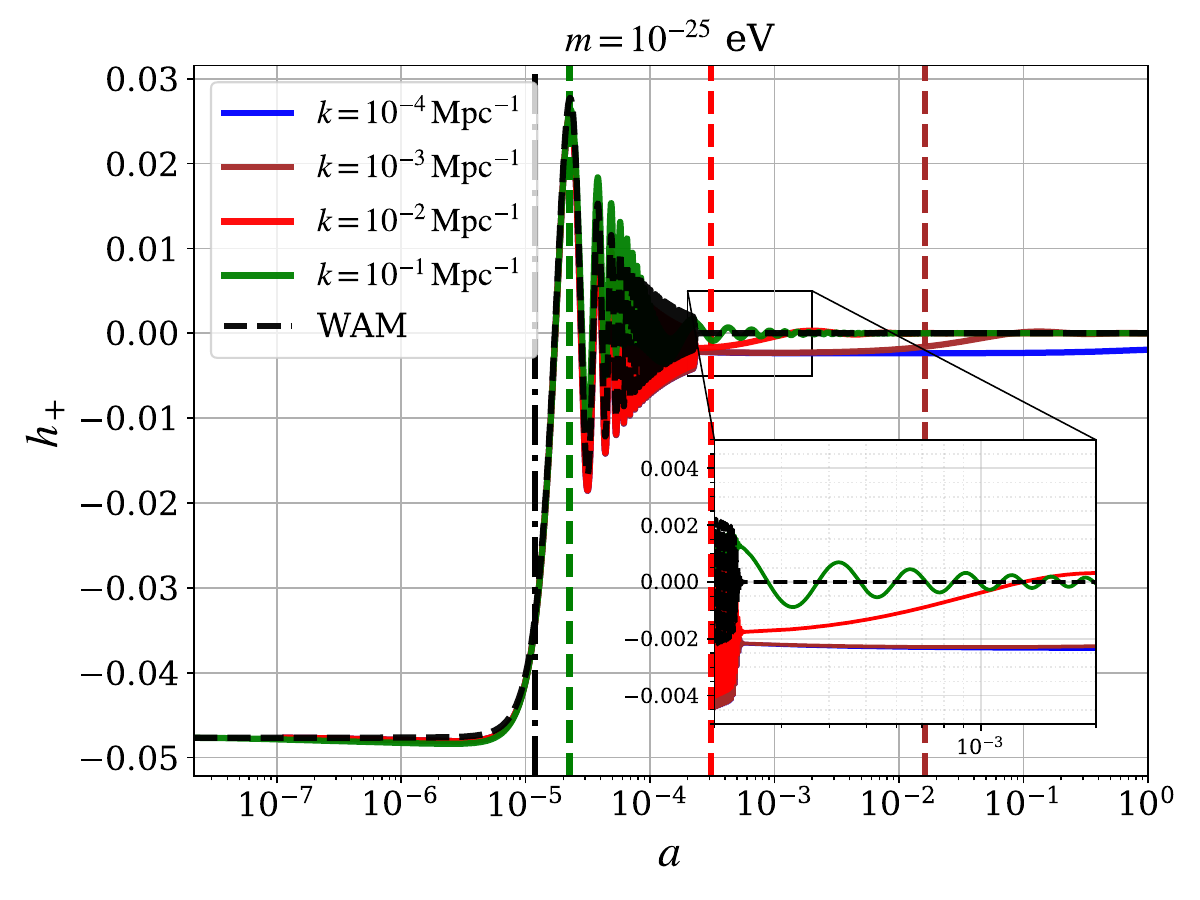}
    \caption{Numerical solution for the tensor perturbation amplitude, for the transversal Fourier modes ($\gamma_k = \pi/2$), and for $m=10^{-22}$eV (left panel) and $m=10^{-25}$eV (right panel). We assume that the vector field is the only source of the tensor perturbations. The dashed colored vertical line indicates the horizon entry $k = \Hconf (a_*)$, and the black dash-dot vertical line the time of oscillation $m = H(a_{\rm osc})$. The black dashed curve corresponds to the Weinberg adiabatic mode for each mass, given by Eq. \eqref{WAM_h_+}. We set the initial conditions such that $w_+=0$.}
\label{fig:h_+}
\end{figure*}

In Fig. \ref{fig:h_+} we show the numerical solutions of Eq. (\ref{einstein_h_plus}) for transversal Fourier modes ($\gamma_k = \pi/2$), for $m=10^{-22}$eV (left panel) and $m=10^{-25}$eV (right panel). 
The dash-dot vertical line indicate the time of oscillation defined as $m = H(a_{\rm osc})$. 
The dashed vertical lines indicate horizon entry for each mode, i.e. $k = \Hconf(a_*)$. 
We set the initial conditions such that $w_+=0$, so the amplitude goes to zero after the field starts oscillating, at leading order in the metric shear.
We can see that before the mode enters the horizon, the tensor perturbation is approximately given by the Weinberg adiabatic mode solution (the black dashed curve). As expected, the approximation improves for larger masses, since the metric shear decreases with increasing mass.
After the mode enters the horizon, we can see the usual decay of the tensor amplitude.
For the lightest masses, $m\sim10^{-25}$ eV, we can see a discrepancy between the analytic Weinberg mode and the numerical solution (see the zoom from the right panel of Fig. \ref{fig:h_+}). This discrepancy is caused by next-to-leading order effects in the metric shear, which we neglect here for simplicity. We leave the implementation of such corrections for future work.

Regarding the evolution of $h_{\times}$, equation \eqref{WAM_h_+} shows that, before the mode enters the horizon, the evolution is given by $\sigma_{\times}$. However, since the background field has no components along the $\hat{e}_2$ direction by construction, $\sigma_{\times} = 0$. Therefore, by setting $w_{\times} = 0$, we have that $h_{\times} = 0$ for the whole evolution. This behavior is confirmed by our numerical results, although we do not show it for simplicity, as the amplitude remains negligible at all times.

\begin{figure*}[th]
    \centering
    \includegraphics[width=0.8\linewidth]{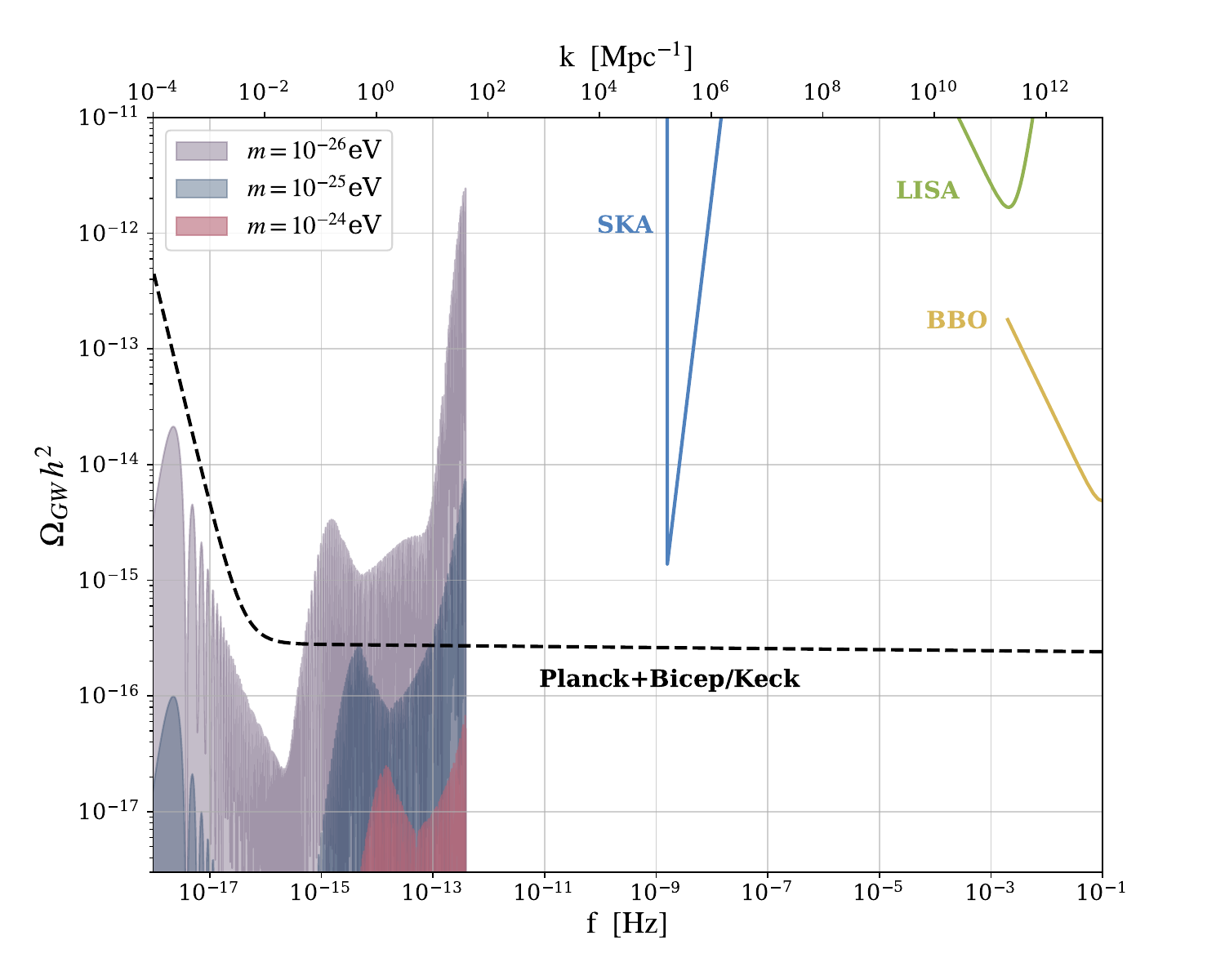}
    \caption{Gravitational wave abundance at present time (shadowed regions), calculated as defined in Eq. \eqref{GW_abundance_def}, for the gravitational waves generated from the transversal mode ($\gamma_k = \pi/2$). The curves correspond to the lowest value each detector is sensitive to. For the current detectors, the gravitational wave abundance from the VFDM is not observable. The dashed black curve indicates the gravitational wave spectrum generated by single-field inflation using $r=0.036$, parameterized by Eq. \eqref{parameterization_cmb_constraint}.}
    \label{fig:GW_spectrum}
\end{figure*}

In Fig. \ref{fig:GW_spectrum} we present the gravitational wave abundance at present-day for different masses of the vector field and for $\gamma_k = \pi/2$. We note that the spectrum is anisotropic, and its amplitude is multiplied by a factor of $\sin^4(\gamma_k)$.
For the largest wavenumbers we find numerical limitations. The spectrum is plotted until the maximum wave-number \texttt{class.VFDM} can integrate. 
We also plot the sensitivities of several gravitational wave detectors \cite{Moore:2014lga}. 
The black dashed curve corresponds to the maximum abundance of gravitational waves allowed by combined Planck+Bicep observations \cite{BICEP:2021xfz}, which can be parameterized as \cite{Clarke:2020bil}
\begin{equation}
    \Omega_{\rm GW}(k)h^2 = \frac{3}{128}\Omega_{r,0} h^2 \mathcal{P}_T(k) \left[\frac{1}{2}\left(\frac{k_{\rm eq}}{k}\right)^2 + \frac{16}{9}\right]\,,
    \label{parameterization_cmb_constraint}
\end{equation}
where $k_{\rm eq} = \sqrt{2} H_0 \Omega_{m,0} / \sqrt{\Omega_{r,0}}\,$, with $H_0$ the Hubble parameter at present time, $\Omega_{r,0}$ and $\Omega_{m,0}$ the radiation and matter abundance at present time respectively; and where we used the consistency relation for slow roll models, i.e. $n_t = -r/8$.
A gravitational wave background at the moment of last scattering produces anisotropies in the CMB through the Sachs-Wolfe effect \cite{Turner:1993vb, Maggiore:1999vm}. 

As mentioned before, we set $\mathcal{A}_{+}=\mathcal{A}_{+s}=0$. Therefore, for non-relativistic modes there is a negligible production of gravitational waves. 
In the numerical result, the first peak is due to the presence of a non-vanishing tensor perturbation $h_+$ outside the horizon which becomes more important for the lightest masses. The evolution of $h_+$  can be seen in the zoom in Fig. \ref{fig:h_+}. As mentioned  above, the numerical solution is different from the Weinberg solution  (which approaches to a constant after averaging over the field oscillations) and the difference is an effect of second order in $|\sigma_{ij}|$. Once the mode enters the horizon, it evolves as in the $\Lambda$CDM scenario (see the zoom in Fig. \ref{fig:h_+}).
The second peak corresponds to the Jeans scale at oscillation ($k_J = a_{\rm osc} \sqrt{m H_{\rm osc}}$) of each mass. 
Finally, the increase in the spectrum at large frequencies is given by the evolution of relativistic modes. We note, however, that on such small scales it becomes necessary to include second-order corrections in cosmological perturbation theory. Although the gravitational waves can still be consistently described at linear order, the stress–energy tensor of the vector field receives small-scale contributions that must be taken into account.   

\section{Conclusions}\label{sec:Conclusions_results}

In this paper we studied the production of gravitational waves due to the evolution of an ultralight vector field dark matter. 
This model presents the challenge of being anisotropic even at the background level, producing a mixing between the SVT sectors. This mixing affects the evolution of the tensor perturbations.

We showed that Einstein equations in the presence of the VFDM can be approximately solved outside the horizon using the Weinberg construction of the adiabatic mode (at leading order in the anisotropies, which are perturbatively small at early times for the masses considered in the paper). 

Finally, we presented an implementation  of the tensor modes in \texttt{class.VFDM}, our modified version of the Einstein-Boltzmann solver \texttt{CLASS} \cite{blas2011cosmic}, extending the scalar implementation of Ref. \cite{Chase:2024wsq}. In Fig. \ref{fig:GW_spectrum} we show the gravitational wave abundance at present time for the transversal Fourier modes (that is, by setting $\gamma_k = \pi/2$), for tensor modes produced during radiation era and afterwards. We emphasize that the spectrum is anisotropic. For a general orientation of the Fourier mode, it is multiplied by a factor $\sin^4(\gamma_k)$.

We noticed that a consistent study of the backreaction of tensor modes on the scalar sector would require a further extension of the code we present here. The reason is that the estimated effect of the tensor modes turns out to be  of the same order as terms in the VFDM stress tensor that are linear in the metric shear, which we are neglecting here. As for future work, it would be worth computing such corrections that are linear in the metric shear, to improve the precision of the numerical implementation, and to study the impact of the SVT mixing in cosmological observables such as the CMB multipoles with \texttt{class.VFDM}.

Furthermore, it would be useful to study a model of anisotropic inflation that generates a coherent vector field, such as the one considered in this work, from which the correlations at the end of inflation can be computed.

\begin{acknowledgments} 
We thank Nahuel Mirón Granese for useful discussions.
This work has been supported by CONICET and UBA. We acknowledge the use of the xAct - xPand package for Mathematica \cite{Pitrou:2013hga, Brizuela:2008ra}.
\end{acknowledgments}

\appendix

\section{Background evolution}\label{appendix:background_equations}

In this section we review the background evolution of the VFDM model following Ref. \cite{Chase:2023puj}. We start by considering the homogeneous part of the vector equation of motion, Eq. (\ref{eq_motion_vector}). This equation can be split into a constraint equation for the temporal component $A_0$, plus the equations of motion for the spatial components $A_i$, obtaining   
\begin{equation}
    A_0 = 0, \qquad A_i^{\prime\prime} - 2 {\sigma^k}_i\,A_k^{\prime} + m^2a^2 A_i = 0\,.
    \label{vector_background_eq}
\end{equation}

The background fluid variables of the vector field in Bianchi I are
\begin{align}
    \rho_A &= \frac{1}{2 a^4} \left[ A_{i}^{\prime}A_{j}^{\prime} + m^2 a^2 A_{i}A_{j} \right]\,\gamma^{ij}\,, \label{rho_A_background}\\[5pt]
    P_A &= \frac{1}{6 a^4} \left[ A_{i}^{\prime}A_{j}^{\prime} - m^2 a^2 A_{i}A_{j} \right] \,\gamma^{ij}\,, \label{background_pressure_expression}\\[5pt]
    {\Sigma_A^i}_j &= \frac{1}{a^4} \bigg[ \frac{1}{3}A_{k}^{\prime} A_{l}^{\prime}\, \gamma^{kl} {\gamma^i}_j - A_{k}^{\prime} A_{j}^{\prime} \, \gamma^{ik}  \\[3pt]
    &\qquad+ m^2 a^2 (A^i A_j - \frac{1}{3}A^k A_k {\gamma^i}_j)  \bigg]\,, \nonumber 
\end{align}
where ${\Sigma_A^i}_j$ is the vector anisotropic stress (i.e., the traceless part of the energy-momentum tensor). In what follows we use the convention of lowering and raising latin indices with the ``spatial metric" $\gamma_{ij}$ and its inverse $\gamma^{ij}$ respectively. The vector field is the only species  we consider here  that has anisotropies at the background level. Then, Einstein's non-diagonal equations are only sourced by the vector field,
\begin{equation}
    ({\sigma^i}_j)^{\prime} + 2 \Hconf\, {\sigma^i}_j = \frac{a^2}{m_P^2} {\Sigma_A^i}_j\,,
    \label{metric_shear_eq}
\end{equation} 
with $m_P$ the Planck mass.

Now we solve the background equations perturbatively in the metric shear. At early times, Eq. (\ref{vector_background_eq}) has a growing mode solution at $\mathcal{O}(m/H)$, given by $\vec{A}=c\tau\hat{A}$, where $c$ is an integration constant, and $\hat{A}$ is fixed in space at leading order. With the solution of the vector field, we can calculate its corresponding fluid variables, obtaining
\begin{align}
    \rho_A &= 3P_A = \frac{c^2}{2a^4}\,, \label{cond_ini_background_fluid_1}\\
    {\Sigma_A^i}_j &= 6 P_A  (\hat{A}_i\hat{A}_j - \frac{{\delta^i}_j}{3})w\,.
    \label{cond_ini_background_fluid_2}
\end{align}
Now we can calculate the constant $c$ in terms of the cosmological parameters and the mass of the field, as 
\begin{equation}
    c^2\simeq 2\, \rho_{cr,0} \,\Omega_{\rm DM,0}\, \Omega_{r,0}^{1/4}\left(\frac{H_0}{m}\right)^{1/2} \,.
    \label{constant_vector_ini}
\end{equation}

With the vector anisotropic stress at hand, we can now solve (\ref{metric_shear_eq}) for the shear tensor, obtaining Eq. (\ref{sigmaT}). Finally, by projecting this solution using $\{\hat{e}_1, \hat{e}_2, \hat{k}\}$, we obtain the expressions for the different components of the shear tensor,
\begin{subequations}
\begin{align}
    \sigma_{\parallel} &= \frac{-2 c_L^2 + c_{t_1}^2}{3 m_P^2 a^2 \Hconf}\,, \label{sigmaparalel}\\[3pt]
    \sigma_{v_1} &= \frac{c_L c_{t_1}}{m_P^2 a^2 \Hconf}\,, \\[3pt]
    \sigma_{+} &= - \frac{c_{t_1}^2}{2 m_P^2 a^2 \Hconf}\,,\label{shear_tensor_early_times}
\end{align}
\label{cond_ini_shear}
\end{subequations}
where $c_L = \cos(\gamma_k) c$ and $c_{t_1} = \sin(\gamma_k) c$.
By construction, the background field does not have components in the $\hat{e}_2$ direction.
Therefore, the remaining components of the shear are not sourced by the vector field, so we set $\sigma_{v_2} = \sigma_{\times} = 0$. 

\section{Synchronous gauge} \label{appendix:synch_gauge}

In this Appendix we present an analysis analogous to that performed in Sec. \ref{sec:SVT_dec_early_times}, now in  the Synchronous gauge. 
The metric perturbations in Synchronous gauge, as defined in Appendix A of Ref. \cite{Chase:2023puj}, read (in Fourier space)
\begin{subequations}
\begin{align}
   \delta g_{00} &= \delta g_{0i} = 0\,,\\[7pt]
   \delta g_{ij} &= a^2\Big[\hat{k}_i\hat{k}_j (h+6\eta) - 2\frac{\sigma_{ij}}{\Hconf} \eta \\[3pt]
   &\qquad+ 2i\, k_{(i} E_{j)} + 2 E_{ij} \Big]\,.\nonumber
\end{align}    
\end{subequations}

We parameterize the tensor degrees of freedom in terms of the usual scalar functions $h_+$ and $h_{\times}$ as in Newtonian gauge (see Eq. \ref{tensor_decomposition}). Also, we work at linear order in any component of $\sigma_{ij}$.

From the temporal component we have the constraint equation
\begin{align}
    \delta A_0 &= \frac{-i\,k}{m^2 a^2 + k^2} \, \big[\delta A_L^{\prime} - \frac{1}{2}A_L^{\prime}(h + 8 \eta )+ i k A_{t_1}^{\prime} E_1 \big]\,.  
\end{align}

In Synchronous gauge, the sources of the equations of motion of the vector field (Eqs.~\ref{eq_mov_delta_A}) are given by
\begin{subequations}
\begin{align}
    S_L &= \frac{1}{2} A_L^{\prime} \left(h^{\prime} + 8\eta^{\prime} \right)+ i k\, A_{t_1}^{\prime} E^{\prime}_1 \,, \\[5pt]
    S_{t_1} &= -\frac{1}{2}A_t^{\prime}(h^{\prime} + 4 \eta^{\prime}) - ik\,  A_{L}^{\prime}  E^{\prime}_1 +2 A_{t_1}^{\prime} h_+ \,.   
\end{align}
\label{sources_synch}
\end{subequations}

We can obtain a solution for the vector field outside the horizon by performing  the corresponding gauge transformation from Newtonian gauge on the Weinberg adiabatic mode of Eq. (\ref{vector_cond_ini}), thus obtaining
\begin{equation}
    \delta \vec{A} =   \left(2 c_L\, \hat{k} - c_{t_1}\,\hat{e}_1 \right) \tau\,\eta_0 + c_{t_1}\tau \left(h_+\,\hat{e}_1 + h_{\times}\,\hat{e}_2\right)\,.
    \label{cond_ini_vec_pert_synch}
\end{equation}
We can check that the previous expression is a solution to the vector equations of motion at early times. With this result, one can show that the adiabatic mode is a solution of Einstein equations outside the horizon, as in Newtonian gauge.

\subsection{Scalar sector}

Now we present the scalar sector of Einstein's equations in Bianchi  I background with VFDM.  The Einstein tensor at linear order in the metric shear can be written as
\begin{subequations} 
    \begin{align}
        \delta {G^0}_0 &= 2k^2 \eta - \Hconf h^{\prime} + 3 \sigma_{\parallel} \eta^{\prime} + 2 \sigma_+ h_+^{\prime} \label{einstein_00_synch}\,,\\[3pt]
        &\quad + i \sigma_{v_1} E_1 \nonumber \\[5pt]
        \hat{k}^i \delta {G^0}_i &= 2k^2 \eta^{\prime} - 3 k^2 \sigma_{\parallel} \eta - k^2 \sigma_+ h_+  \label{einstein_0i_synch}\,,\\[5pt]
        \delta {G^i}_i &= 3 h^{\prime\prime} + 6 \Hconf h^{\prime} - 6 k^2 \eta + 18 \sigma_+ h_+^{\prime} \label{einstein_ii_synch} \\[3pt]
        &\quad+ 9 i k \sigma_{v_1} E_1^{\prime} \,,\nonumber\\[5pt]
        \delta {G}_{\parallel} &= 3 h^{\prime\prime} + 18 \eta^{\prime\prime} + 6 \Hconf (h^{\prime} + 6 \eta^{\prime})  \label{einstein_ij_synch}\\[3pt]
        &\quad+ 18 i \sigma_{v_1} E_1^{\prime} + 3 i k(\sigma_{v_1}^{\prime} + 2 \Hconf \sigma_{v_1}) E_1  \,, \nonumber
    \end{align}
    \label{einstein_eqs_synch}
\end{subequations}

The fluid variables on the rhs are defined as in Eq. (\ref{fluid_variables_definitions}), and are given by
\begin{subequations} 
\begin{align}
    \delta \rho_A  = &\frac{1}{a^4} \left[A^{\prime}_L \left(\delta A^{\prime}_L - i\,k\,\delta A_0\right) +  A_{t_1}^{\prime} \delta A_{t_1}^{\prime} \right.\label{delta_rho_synch} \\[4pt]
    & \left. + m^2 a^2 (A_L \,\delta A_L + A_{t_1} \,\delta A_{t_1})\right] \nonumber\\[4pt]
    &+2 \sin(\gamma_k)^2\rho_A (\eta - h_+) - \cos(\gamma_k)^2 \rho_A ( h + 4 \eta )\,,\nonumber
    \end{align}
\begin{align} 
    \delta P_A = &\frac{1}{3a^4} \left[A^{\prime}_L (\delta A^{\prime}_L- i\,k\,\delta A_0)  + A_{t_1}^{\prime}  \delta A_{t_1}^{\prime} \right. \label{delta_P_synch}   \\[4pt] 
    &\left.- m^2 a^2 (A_L \,\delta A_L + A_{t_1} \,\delta A_{t_1}) \right]\nonumber \\[4pt]
    &+2 \sin(\gamma_k)^2 P_A (\eta - h_+) - \cos(\gamma_k)^2 P_A ( h + 4 \eta )\,,\nonumber  \nonumber 
\end{align}
\begin{align} 
    (\rho_A &+ P_A) \theta_A = \frac{k^2}{a^4} A^{\prime}_T \delta A_{T_1} -i \frac{m^2 \, k}{a^2} \, A_L\, \delta A_0\,, 
    \end{align}
\begin{align} 
    (\rho_A &+ P_A) \delta\Sigma_{A\parallel}  = \frac{4}{3 a^4} \left[A^{\prime}_L (\delta A^{\prime}_L - i\,k\,\delta A_0)  \right.\label{delta_shear_synch}\\[4pt]
    &\left.+ m^2 a^2 (A_{t_1} \,\delta A_{t_1} - A_L \,\delta A_L)  - \frac{1}{2}A_{t_1}^{\prime}\delta A^{\prime}_{t_1} \right] \nonumber \\[4pt]
    &- 4 \cos(\gamma_k)^2 P_A (h + 4 \eta) - 4 \sin(\gamma_k)^2 P_A \, (\eta - h_+) \nonumber\,.
\end{align}
\label{fluid_variables_synchronous}
\end{subequations}

Now we will study the solutions of the system outside   horizon. These solutions can be obtained either by solving the equations in the $k\ll H$ limit, and by looking for the attractor solutions; or by performing a gauge transformation on the solutions found in Eqs. (\ref{adiabatic_conds_ini}). By doing this, we obtain for the metric and radiation perturbations \cite{Ma:1995ey}
\begin{subequations}
\begin{align}
    \eta &= \eta_0 - \alpha \frac{\eta_0}{2} (k \tau)^2\,,\label{ecetaLCDM}\\[3pt]
    h &= \frac{\eta_0}{2} (k \tau)^2\,, \label{h_scalar_ini}\\[3pt]
    \delta_{\gamma} &= - \frac{\eta_0}{3} (k\tau)^2 \,, \label{delta_gamma_ini}
\end{align}
\label{adiabatic_cond_ini_synch}
\end{subequations}
where $\alpha$ is a constant that is fixed with Eq.~(\ref{einstein_0i_synch}) (in $\Lambda$CDM  $\alpha= (5 + 4 R_{\nu})(15 + 4 R_{\nu})^{-1}/12$, with $R_{\nu}$ the ratio of neutrinos to radiation energy density \cite{Ma:1995ey}), and $\eta_0$ is an integration constant.
The initial conditions for the rest of the species are calculated by imposing the adiabatic initial condition (Eq. \ref{adiabatic_conds_ini}).

We see that the tensor perturbation $h_+$ enters in the scalar sector of the Einstein tensor explicitly. However, as in Newtonian gauge, using that $h_+=const$ for the adiabatic mode when $H\gg m$, and that the metric perturbations $\eta$ is constant and $h$ vanishes in the $k\to0$ limit, we see that the terms containing tensor perturbations in Eqs. (\ref{einstein_00_synch}), (\ref{einstein_ii_synch}) and (\ref{einstein_ij_synch}) are subleading. Therefore, the mixing in  such equations is negligible at leading order in the anisotropies and outside the horizon. 

Now we focus on the stress-energy of the vector field. As can be seen by inserting the adiabatic solution (\ref{cond_ini_vec_pert_synch}) in Eqs. (\ref{delta_rho_synch}), (\ref{delta_P_synch}) and (\ref{delta_shear_synch}), the mixing becomes negligible after a cancellation between the vector field and the tensor metric perturbations.
However, we cannot neglect the metric shear in Eq. (\ref{einstein_0i_synch})  outside the horizon, since in the $k\to0$ limit $\eta$ becomes constant, so the Bianchi correction dominates the left-hand side of the equation.
By inserting the initial conditions in Synchronous gauge (Eqs. (\ref{cond_ini_vec_pert_synch}) and (\ref{adiabatic_cond_ini_synch})) in Eq. (\ref{einstein_0i_synch}), we can see that the terms containing the shear tensor on the left-hand side are canceled by $(\rho+P)\theta_A$.
Therefore, to account for this cancellation in the numerical code, we include the Bianchi terms in Einstein $0i$ in \texttt{class.VFDM}. We therefore have that, as in Newtonian gauge, despite the mixing cannot be neglected outside the horizon for this equation, the system can still be solved using the Weinberg construction. 

\subsection{Vector sector} 

We present the vector sector in Synchronous gauge for completeness, although it is not implemented in \texttt{class.VFDM}.
For the vector sector we consider that the VFDM is the only species that sources the vector modes. 
As in Newtonian gauge, the Einstein vector equations can be obtained by projecting $\delta {G^0}_i$ in the directions $\hat{e}_1$ and $\hat{e}_2$, thus obtaining respectively
\begin{align}
    &\quad\frac{k^2}{2} \left[E^{\prime}_1 + (\frac{\sigma_{\parallel}}{2}-\sigma_+)E_1\right] + i \frac{k}{2} \sigma_{v_1} h + i k \frac{\sigma_{v_1}}{\Hconf} \eta^{\prime}\label{einstein_vec_synch}\\
    & + i k \left[(6-\frac{\Hconf^{\prime}}{\Hconf^2})\sigma_{v_1}+\frac{\sigma_{v_1}^{\prime}}{\Hconf}\right] \eta   - 2 i k \sigma_{v_1} h_+ =  \nonumber\\
    &\qquad -\frac{1}{m_P^2} \left[ m^2 A_{t_1} \delta A_0  - i \frac{k}{a^2} A_L^{\prime}\delta A_{t_1}  \right]\,, \nonumber
\end{align}
\begin{align}
    &\quad\frac{k}{2} \left[E^{\prime}_2 + (\frac{\sigma_{\parallel}}{2}-\sigma_+)E_2\right] - 2 i \sigma_{v_1} h_{\times} = \frac{i}{m_P^2 a^2} A_L^{\prime}\delta A_{t_2} \,. 
\end{align}

\subsection{Tensor sector}

The tensor perturbations are gauge invariant, so their equations of motion in Synchronous gauge are still given by Eqs. (\ref{einstein_eqs_tensor_newt}). The sources $H_+$ and $A_+$, however, need to be written in terms of the variables in Synchronous gauge, as
\begin{align}
    &H_+ = \frac{\sigma_+}{\Hconf} \eta^{\prime\prime} + \frac{\sigma_+}{2} h^{\prime} - 2\left[(1 - \frac{\Hconf^{\prime}}{\Hconf^2}) \sigma_+ + \frac{\sigma_+^{\prime}}{\Hconf}\right] \eta^{\prime} \label{einstein_tensor_source_H_synch}\\
    &+ \bigg[\Hconf^{-1}(\sigma_+^{\prime\prime} + (k^2 - 2 \Hconf^{\prime})\sigma_+)  - \Hconf^{-2} (\Hconf^{\prime\prime} \sigma_+ + 2 \Hconf^{\prime} \sigma_+^{\prime}) \nonumber\\
    &+ 2 \sigma_+^{\prime} + 2 \frac{\Hconf^{\prime\, 2}}{\Hconf^3} \sigma_+\bigg] \eta + i k \sigma_{v_1} E_1^{\prime}  + \frac{i k}{2} (\sigma_{v_1}^{\prime} + 2 \Hconf \sigma_{v_1}) E_1 \nonumber \,,
\end{align}
\begin{align}
    A_+ &= - A_{t_1}^{\prime} \delta A_{t_1}^{\prime} + m^2 a^2 A_{t_1} \delta A_{t_1} - \left[ A_{t_1}^{\prime\,2} \right.\label{einstein_tensor_source_A_synch}\\[3pt]
    &\left. - m^2 a^2 A_{t_1}^2 \right]\eta + \frac{i k}{2} \left[ A_L^{\prime} A_{t_1}^{\prime} - m^2 a^2 A_L  A_{t_1}) \right] E_1 \nonumber \,.
\end{align}

\section{Implementation in \texttt{class.VFDM}} \label{appendix:class_implementation}

\subsection{Background}

In this appendix, we describe the implementation of the VFDM model in \texttt{class.VFDM}, including the mixing of the scalar and tensor sector. This implementation is performed in Synchronous gauge, and is based on Ref. \cite{Chase:2024wsq}. In here, we present the equations that need to be modified to include the tensor sector. We start by reviewing the background implementation. To solve the background equations, we use the following change of variables for the vector field, 
\begin{align}
    \vec{A} &= \sqrt{6}\, m_P \frac{\Hconf}{m} \sqrt{\Omega_A} \sin \left(\frac{\theta}{2}\right) \hat{A}\,, \\
    \vec{A}^{\,\prime} &= \sqrt{6}\, m_P \, a\Hconf \sqrt{\Omega_A} \cos \left(\frac{\theta}{2}\right) \hat{A\,.}  
\end{align}
In this way, we use the vector abundance $\Omega_A$ as a dynamical variable, and also parameterize the vector oscillations at late times in the new function $\theta$. With this parameterization, we obtain that the equation of state for the field is now given by $w_A = \frac{1}{3}\cos(\theta)$. Furthermore, the background equations can be written as  
\begin{subequations}
\begin{align}
    \dot{\theta} &= \sin(\theta)+y\,, \\[5pt]
    \dot{\Omega}_{A} &= 3(w_T - w_A) \Omega_{A} \,, \\[5pt]
    \dot{y} &= \frac{3}{2}(1+w_T)y\,,
\end{align}\label{background_system_of_eqs}
\end{subequations}
where a dot derivative indicates $\frac{d}{d \log a}$ and $y =2  {m a}/{\Hconf}$. The previous equations are the ones implemented in the \textit{background} module of \texttt{class.VFDM}.

The background equations can be solved analytically at early and late times. For the first, we obtain that $\theta \propto a^2 \ll 1$ while $\Omega_A \sim const$, so $w_A \sim 1/3$ and thus the vector field acts as an anisotropic radiation component. At late times, we obtain that $\theta \sim 2mt$, so in the regime of $m\gg H$ we can average the equations on cosmological timescales, thus obtaining $\langle w_A\rangle\, \sim\, \langle\cos(\theta)\rangle = 0$, where $\langle \dots\rangle$ indicates average over cosmological timescales, thus recovering a CDM equation of state. This time-average is performed in \texttt{class.VFDM} by replacing each trigonometric function containing $\theta$, by a function that smoothly vanishes, implemented as
\begin{align}
    \sin_*(\theta) &\to \frac{1}{2} \left[1 - \tanh\left(\theta_{tol}(\theta^2 - \theta_{th}^2)\right)\right]\sin(\theta)\,, \\[3pt]
    \cos_*(\theta) &\to \frac{1}{2} \left[1 - \tanh\left(\theta_{tol}(\theta^2 - \theta_{th}^2)\right)\right]\cos(\theta)\,.
\end{align}
For the results presented here, we used $\theta_{tol} = 0.1$ and $\theta_{th} = 100$.

\subsection{Linear perturbations}

Now we present the modifications carried out in the linear regime to include the tensor perturbations. The tensor perturbations are only affected by the $\hat{e}_1$ components of the vector field. We therefore focus on the implementation of the transversal Fourier modes, since the longitudinal ones where unmodified with respect to \cite{Chase:2024wsq}.
We start by performing a change of variables, and write the vector field in terms of two new variables $\{\delta_{i,0}, \delta_{i,1}\}$, where $i = L, T$. For the Fourier longitudinal components we found it convenient to perform the following change of variables,
\begin{subequations}
\begin{align}
    \delta A_L =& \frac{1}{2} \left[A_L \, (\delta_{L,0} + h + 8 \eta) + \frac{1}{m a} A_L^{\prime} \,\delta_{L,1}\right], \\[4pt]
    \delta A_L^{\prime} =& \frac{\omega}{2} \left[A_L^{\prime} \, (\delta_{L,0} + h + 8 \eta) - m a A_L \,\delta_{L,1}\right]\,.
\end{align}
\end{subequations}
For the transversal Fourier components, we performed the change of variables
\begin{subequations}
\begin{align}
    \delta A_T =& \frac{1}{2} \left[A_T \, \delta_{T,0} - \frac{1}{m a} A_T^{\prime} \,\delta_{T,1}\right], \\[4pt]
    \delta A_T^{\prime} =& \frac{\omega}{2} \left[A_T^{\prime} \, \delta_{T,0} - m a A_T  \,\delta_{T,1}\right],
\end{align}
\label{change_of_variables_AT}
\end{subequations}
with $\omega = \sqrt{1 + k^2/ma}$. We used the same change of variables from Eq. \eqref{change_of_variables_AT} for the $\hat{e}_1$ and $\hat{e}_2$ components of the vector field perturbations, so $T = t_1, t_2$.

Now we present the equations of motion of the vector field in these new variables. As mentioned before, the coupling with the tensor sector is given by the $\hat{e}_1$ component of the vector field. We thus present the modified equations for $\{\delta_{T,0}, \delta_{T,1}\}$,
\begin{widetext}
\begin{align}
    &\dot{\delta}_{t_1,0} \;=\;\frac{1}{4\,y\,\omega^2}\,
    \Big[y\,\omega \big(4\sin\theta\,\omega + y(\omega-1)\big(1+\cos\theta + (\cos\theta-1)\omega\big)\big) -4\kappa \big(2\sin\theta + (1+\cos\theta)\,y\,\omega\big) \Big] \delta_{t_1,1} \label{eq_delta1_0}\\[6pt]
    &\quad+ \frac{1}{4\,y\,\omega^2}\, \Big[8\kappa(1+\cos\theta)
    + \sin\theta\,y\,\omega\big(y(\omega^2-1) - 4\kappa \big)
    \Big] \delta_{t_1,0} - \frac{1+\cos\theta}{2\omega}\,\dot{h} - \frac{2(1+\cos\theta)}{\omega}\,\dot{\eta}
    + \frac{2(1+\cos\theta)}{\omega}\,\dot{h}_+ \,, \nonumber \\[10pt]
    &\dot{\delta}_{t_1,1} \;=\; \frac{1}{4\,y\,\omega^2}\, \Big[ y^2(\omega-1)\omega \big(\omega-1 + \cos\theta(1+\omega)\big) -4\kappa \big(2\sin\theta + (\cos\theta-1)\,y\,\omega\big) \Big] \delta_{t_1,0} \label{eq_delta1_1}\\[6pt]
    &\quad+ \frac{1}{4\,y\,\omega^2}\,
    \Big[\kappa\big(8 - 8\cos\theta + 4\sin\theta\,y\,\omega\big)
    + y\,\omega\big(4\cos\theta\,\omega- \sin\theta\,y(\omega^2-1)\big)\Big] \delta_{t_1,1} + \frac{\sin\theta}{2\omega}\,\dot{h} + \frac{2\sin\theta}{\omega}\,\dot{\eta} - \frac{2\sin\theta}{\omega}\,\dot{h}_+ \,, \nonumber\\[10pt]
    &\dot{\delta}_{t_2,0} \;=\;\frac{1}{4\,y\,\omega^2}\,
    \Big[y\,\omega \big(4\sin\theta\,\omega + y(\omega-1)\big(1+\cos\theta + (\cos\theta-1)\omega\big)\big) -4\kappa \big(2\sin\theta + (1+\cos\theta)\,y\,\omega\big) \Big] \delta_{t_2,1} \label{eq_delta2_0}\\[6pt]
    &\qquad+ \frac{1}{4\,y\,\omega^2}\, \Big[8\kappa(1+\cos\theta)
    + \sin\theta\,y\,\omega\big(y(\omega^2-1) - 4\kappa \big)
    \Big] \delta_{t_2,0} 
    + \frac{2(1+\cos\theta)}{\omega}\,\dot{h}_{\times} \,, \nonumber \\[10pt]
    &\dot{\delta}_{t_2,1} \;=\; \frac{1}{4\,y\,\omega^2}\, \Big[ y^2(\omega-1)\omega \big(\omega-1 + \cos\theta(1+\omega)\big) -4\kappa \big(2\sin\theta + (\cos\theta-1)\,y\,\omega\big) \Big] \delta_{t_2,0} \label{eq_delta2_1}\\[6pt]
    &\qquad+ \frac{1}{4\,y\,\omega^2}\,
    \Big[\kappa\big(8 - 8\cos\theta + 4\sin\theta\,y\,\omega\big)
    + y\,\omega\big(4\cos\theta\,\omega- \sin\theta\,y(\omega^2-1)\big)\Big] \delta_{t_2,1} - \frac{2\sin\theta}{\omega}\,\dot{h}_{\times} \,, \nonumber
\end{align}
\end{widetext}
where $\kappa =  k^2/(2 m a \Hconf)$. We thus see that the dynamics of the vector field is coupled to the tensor sector not only before the field starts oscillating, but also afterwards.

Now we present the fluid variables of Eq. (\ref{fluid_variables_synchronous}), written in terms of the new set of variables,
\begin{align}
     \delta \rho_A &= \rho_A \frac{\cos(\gamma_k)^2}{4 \kappa + y} \bigg[(2\,\kappa\, (1-\cos\left(\theta\right))  + y) \delta_{L,0} \label{delta_rho}\\[3pt]
     &+ 2 \sin\left(\theta\right) \,\kappa\, \delta_{L,1}\bigg] + \frac{1}{2} \rho_A \sin(\gamma_k)^2 \nonumber\\[3pt]
     &\times\left[ (1 + \omega + \cos(\theta)(\omega-1)) \delta_{t_1,0} - \sin(\theta) (\omega-1) \delta_{t_1,1} \right] \nonumber\\[3pt]
     &+ \rho_A\,(3 + \cos(2\gamma_k)) \eta - 2 \rho_A \sin(\gamma_k)^2 h_+\,, \nonumber
\end{align}
\begin{align}
    \delta P_A &= \frac{\rho_A}{3}\frac{\cos(\gamma_k)^2}{4\kappa + y} \bigg[((2\kappa+y)\cos\left(\theta\right)  - 2\kappa)\, \delta_{L,0} \\[3pt]
    &- \sin\left(\theta\right) (2\kappa+y) \delta_{L,1}\bigg] +\frac{\rho_A}{6} \sin(\gamma_k)^2\nonumber\\[3pt]
    &\times\left[(\omega-1+\cos(\theta)(1+\omega))\delta_{t_1,0} - \sin(\theta)(1+\omega) \right]\nonumber\\[3pt]
    &-\sin(\theta)\delta_{t_1,1}] + \frac{\rho_A}{3}\,(3 + \cos(2\gamma_k)) \cos\left(\theta\right) \eta \nonumber\\
    &- \frac{2}{3} \rho_A \cos(\theta)\sin(\gamma_k)^2 h_+ \,,\nonumber
\end{align}
\begin{align}
    (\rho_A + &P_A) \theta_{A} =  \rho_A \cos(\gamma_k)^2\,\frac{ a H y \kappa}{4\kappa+y} \bigg[-\sin(\theta) \delta_{L,0} \\[3pt]
    &+ (1-\cos(\theta)) \delta_{L,1}\bigg] + \rho_A \sin(\gamma_k)^2 a H \kappa \nonumber\\[3pt]
    &\times\left[\sin(\theta)  \delta_{t_1,0} + (1+\cos(\theta))  \,\delta_{t_1,1}\right]\,,\nonumber
\end{align}
\begin{align}
    (\rho_A &+ P_A) \delta\Sigma_{A,\parallel} = \frac{4}{3}\rho_A \cos(\gamma_k)^2 \frac{1}{4\kappa+y} \times\\[3pt]
    &\times \big[(2 \kappa \left(-1+\cos (\theta)\right)  + \cos (\theta) y) \delta_{L,0} \nonumber\\[3pt]
    &- \sin (\theta) \left(2\kappa+ y\right) \delta_{L,1} \big] + \frac{1}{3}\rho_A \sin(\gamma_k)^2  \nonumber\\[3pt]
    &\times\left[(\omega-1+\cos(\theta)(1+\omega)) \delta_{t_1,0} + (1+\omega)\sin(\theta) \delta_{t_1,1}\right] \nonumber\\[3pt]
    &+\frac{2}{3} \rho_A  \cos(\theta) \left[(3 + 5 \cos(2\gamma_k))  \eta + 2 \sin(\gamma_k)^2 h_+\right] \,. \nonumber
\end{align}
The previous quantities are the ones implemented in the code, added as an additional species on Einstein equations. 

Now we calculate the initial conditions including the tensor perturbation. We calculate this by either looking for the attractor solution of Eqs. \eqref{eq_delta1_0}, \eqref{eq_delta1_1}, \eqref{eq_delta2_0} and \eqref{eq_delta2_1} at early times, or by transforming the solutions from Eq. (\ref{cond_ini_vec_pert_synch}). In both ways, we obtain 
\begin{subequations}
\begin{align}
    \delta_{t_1,0} &= \frac{y(2+\kappa y)}{2\kappa + y} (h_+ - \eta_0)\,, \label{cond_ini_T_0}\\[4pt]
    \delta_{t_1,1} &= \frac{2 \kappa y}{2 \kappa + y} (h_+ - \eta_0)\,. \label{cond_ini_T_1} \\[4pt]
    \delta_{t_2,0} &= \frac{y(2+\kappa y)}{2\kappa + y} h_{\times}\,, \label{cond_ini_T2_0}\\[4pt]
    \delta_{t_2,1} &= \frac{2 \kappa y}{2 \kappa + y} h_{\times}\,. \label{cond_ini_T1_1}
\end{align}
\end{subequations}

Finally, we write the equation for the tensor perturbation (\ref{einstein_h_plus}) in these new variables as
\begin{align}
    &h_+^{\prime\prime} + 2 \Hconf h_+^{\prime} + \left[k^2 - \frac{2 a^2 \rho_A}{m_P^2} \sin ^2(\gamma_k) \cos (\theta)\right] h_+  = \nonumber\\[3pt]
    &\quad\frac{a^2 \rho_A}{2m_P^2} \sin^2(\gamma_k) \bigg[ (\omega+1) \sin(\theta) \delta_{t_1,1} - 4 \cos(\theta) \eta \nonumber\\[3pt]
    &\quad+  ((\omega+1) \cos(\theta) + \omega - 1 ) \delta_{t_1,0} \bigg] \nonumber\\
    &+ \left[\sigma_+ (2 - \frac{2 \Hconf^{\prime}}{\Hconf^2}) + 2\frac{\sigma_+}{\Hconf}\right] \eta^{\prime} + \frac{\sigma_+}{\Hconf} \eta^{\prime\prime} - \frac{1}{2} \sigma_+^{\prime} h^{\prime} \,.\label{einstein_tensor_class}
\end{align}

In Sec. \ref{sec:WeinbergAD} we discuss that the initial conditions can be calculated using the construction of the Weinberg adiabatic mode. However, this construction depends on arbitrary constants given in the matrix ${w^{i}}_j$ (see Eq. \eqref{WAM_h_+}). 
In particular, the initial conditions for the gravitational waves depend on a new parameter $w_+$.  To present the results in this paper, we set $w_+=0$. This is achieved by setting initially
\begin{equation}
    h_+ = \frac{\sigma_+}{\Hconf} \frac{\eta_0}{2} \,.
\end{equation}

\bibliographystyle{ieeetr}
\bibliography{reference}

@article{Amin:2022nlh,
    author = "Amin, Mustafa A. and Mirbabayi, Mehrdad",
    title = "{A Lower Bound on Dark Matter Mass}",
    eprint = "2211.09775",
    archivePrefix = "arXiv",
    primaryClass = "hep-ph",
    doi = "10.1103/PhysRevLett.132.221004",
    journal = "Phys. Rev. Lett.",
    volume = "132",
    number = "22",
    pages = "221004",
    year = "2024"
}

@article{Rogers:2020ltq,
    author = "Rogers, Keir K. and Peiris, Hiranya V.",
    title = "{Strong Bound on Canonical Ultralight Axion Dark Matter from the Lyman-Alpha Forest}",
    eprint = "2007.12705",
    archivePrefix = "arXiv",
    primaryClass = "astro-ph.CO",
    doi = "10.1103/PhysRevLett.126.071302",
    journal = "Phys. Rev. Lett.",
    volume = "126",
    number = "7",
    pages = "071302",
    year = "2021"
}

@article{Nakayama:2019rhg,
    author = "Nakayama, Kazunori",
    title = "{Vector Coherent Oscillation Dark Matter}",
    eprint = "1907.06243",
    archivePrefix = "arXiv",
    primaryClass = "hep-ph",
    reportNumber = "UT-19-18",
    doi = "10.1088/1475-7516/2019/10/019",
    journal = "JCAP",
    volume = "10",
    pages = "019",
    year = "2019"
}

@article{Kaneta:2023lki,
    author = "Kaneta, Kunio and Lee, Hye-Sung and Lee, Jiheon and Yi, Jaeok",
    title = "{Misalignment mechanism for a mass-varying vector boson}",
    eprint = "2306.01291",
    archivePrefix = "arXiv",
    primaryClass = "astro-ph.CO",
    doi = "10.1088/1475-7516/2023/09/017",
    journal = "JCAP",
    volume = "09",
    pages = "017",
    year = "2023"
}

@article{Kitajima:2023fun,
    author = "Kitajima, Naoya and Nakayama, Kazunori",
    title = "{Viable vector coherent oscillation dark~matter}",
    eprint = "2303.04287",
    archivePrefix = "arXiv",
    primaryClass = "hep-ph",
    reportNumber = "TU-1181, KEK-QUP-2023-0004",
    doi = "10.1088/1475-7516/2023/07/014",
    journal = "JCAP",
    volume = "07",
    pages = "014",
    year = "2023"
}

@article{Ma:1995ey,
    author = "Ma, Chung-Pei and Bertschinger, Edmund",
    title = "{Cosmological perturbation theory in the synchronous and conformal Newtonian gauges}",
    eprint = "astro-ph/9506072",
    archivePrefix = "arXiv",
    doi = "10.1086/176550",
    journal = "Astrophys. J.",
    volume = "455",
    pages = "7--25",
    year = "1995"
}

@article{Pereira_2007,
	doi = {10.1088/1475-7516/2007/09/006},
	url = {https://doi.org/10.1088%2F1475-7516%2F2007%2F09%2F006},
	year = 2007,
	month = {sep},
	publisher = {{IOP} Publishing},
	volume = {2007},
	number = {09},
	pages = {006--006},
	author = {Thiago S Pereira and Cyril Pitrou and Jean-Philippe Uzan},
	title = {Theory of cosmological perturbations in an anisotropic universe}, 
	journal = {Journal of Cosmology and Astroparticle Physics}
}

@article{Marsh_2016,
	doi = {10.1016/j.physrep.2016.06.005},
	url = {https://doi.org/10.1016%2Fj.physrep.2016.06.005},
	year = 2016,
	month = {jul},
	publisher = {Elsevier {BV}},
	volume = {643},
	pages = {1--79},
	author = {David J.E. Marsh},
	title = {Axion cosmology},
	journal = {Physics Reports}
}

@article{Ferreira_2021,
	doi = {10.1007/s00159-021-00135-6},
	url = {https://doi.org/10.1007\%2Fs00159-021-00135-6},
	year = 2021,
	month = {sep},
	publisher = {Springer Science and Business Media {LLC}},
	volume = {29},
	number = {1},
	author = {Elisa G. M. Ferreira},
	title = {Ultra-light dark matter},
	journal = {The Astronomy and Astrophysics Review}
}

@article{Weinberg_2003,
	doi = {10.1103/physrevd.67.123504},
	url = {https://doi.org/10.1103%2Fphysrevd.67.123504},
	year = 2003,
	month = {jun},
	publisher = {American Physical Society ({APS})},
	volume = {67},
	number = {12},
	author = {Steven Weinberg},
	title = {Adiabatic modes in cosmology},
	journal = {Physical Review D}
}

@article{Planck_2018,
	doi = {10.1051/0004-6361/201833910},
	url = {https://doi.org/10.1051%2F0004-6361%2F201833910},
	year = 2020,
	month = {sep},
	publisher = {{EDP} Sciences},
	volume = {641},
	pages = {A6},
	author = {Planck collaboration},
	title = {Planck 2018 results. VI. Cosmological parameters},
	journal = {Astronomy \& Astrophysics}
}

@article{blas2011cosmic,
  title={The cosmic linear anisotropy solving system (CLASS). Part II: Approximation schemes},
  author={Blas, Diego and Lesgourgues, Julien and Tram, Thomas},
  journal={Journal of Cosmology and Astroparticle Physics},
  volume={2011},
  number={07},
  pages={034},
  year={2011},
  publisher={IOP Publishing}
}

@article{Miravet:2020kuj,
    author = "Miravet, Alfredo D. and Maroto, Antonio L.",
    title = "{Imprint of ultralight vector fields on gravitational wave propagation}",
    eprint = "2012.07505",
    archivePrefix = "arXiv",
    primaryClass = "astro-ph.CO",
    doi = "10.1103/PhysRevD.103.123546",
    journal = "Phys. Rev. D",
    volume = "103",
    number = "12",
    pages = "123546",
    year = "2021"
}

@article{Marsh:2015wka,
    author = "Marsh, David J. E. and Pop, Ana-Roxana",
    title = "{Axion dark matter, solitons and the cusp\textendash{}core problem}",
    eprint = "1502.03456",
    archivePrefix = "arXiv",
    primaryClass = "astro-ph.CO",
    doi = "10.1093/mnras/stv1050",
    journal = "Mon. Not. Roy. Astron. Soc.",
    volume = "451",
    number = "3",
    pages = "2479--2492",
    year = "2015"
}

@article{Chase:2023puj,
    author = "Chase, Tom\'as Ferreira and L\'opez Nacir, Diana",
    title = "{Ultralight vector dark matter, anisotropies, and cosmological adiabatic modes}",
    eprint = "2311.09373",
    archivePrefix = "arXiv",
    primaryClass = "astro-ph.CO",
    doi = "10.1103/PhysRevD.109.083521",
    journal = "Phys. Rev. D",
    volume = "109",
    number = "8",
    pages = "083521",
    year = "2024"
}

@article{Chase:2024wsq,
    author = "Chase, Tom\'as Ferreira and Leizerovich, Mat\'\i{}as and L\'opez Nacir, Diana and Landau, Susana",
    title = "{Cosmological perturbations with ultralight vector dark matter fields: Numerical implementation in class}",
    eprint = "2408.12052",
    archivePrefix = "arXiv",
    primaryClass = "astro-ph.CO",
    doi = "10.1103/PhysRevD.111.103520",
    journal = "Phys. Rev. D",
    volume = "111",
    number = "10",
    pages = "103520",
    year = "2025"
}

@article{Weinberg:tensor_modes,
    author = "Weinberg, Steven",
    title = "{Damping of tensor modes in cosmology}",
    eprint = "astro-ph/0306304",
    archivePrefix = "arXiv",
    reportNumber = "UTTG-02-03",
    doi = "10.1103/PhysRevD.69.023503",
    journal = "Phys. Rev. D",
    volume = "69",
    pages = "023503",
    year = "2004"
}

@article{Pitrou:2013hga,
    author = "Pitrou, Cyril and Roy, Xavier and Umeh, Obinna",
    title = "{xPand: An algorithm for perturbing homogeneous cosmologies}",
    eprint = "1302.6174",
    archivePrefix = "arXiv",
    primaryClass = "astro-ph.CO",
    doi = "10.1088/0264-9381/30/16/165002",
    journal = "Class. Quant. Grav.",
    volume = "30",
    pages = "165002",
    year = "2013"
}

@article{Brizuela:2008ra,
    author = "Brizuela, David and Martin-Garcia, Jose M. and Mena Marugan, Guillermo A.",
    title = "{xPert: Computer algebra for metric perturbation theory}",
    eprint = "0807.0824",
    archivePrefix = "arXiv",
    primaryClass = "gr-qc",
    doi = "10.1007/s10714-009-0773-2",
    journal = "Gen. Rel. Grav.",
    volume = "41",
    pages = "2415--2431",
    year = "2009"
}

@article{Cembranos:2012kk,
    author = "Cembranos, J. A. R. and Hallabrin, C. and Maroto, A. L. and Jareno, S. J. Nunez",
    title = "{Isotropy theorem for cosmological vector fields}",
    eprint = "1203.6221",
    archivePrefix = "arXiv",
    primaryClass = "astro-ph.CO",
    doi = "10.1103/PhysRevD.86.021301",
    journal = "Phys. Rev. D",
    volume = "86",
    pages = "021301",
    year = "2012"
}

@article{Cembranos:2013cba,
    author = "Cembranos, J. A. R. and Maroto, A. L. and N\'u\~nez Jare\~no, S. J.",
    title = "{Isotropy theorem for arbitrary-spin cosmological fields}",
    eprint = "1311.1402",
    archivePrefix = "arXiv",
    primaryClass = "gr-qc",
    doi = "10.1088/1475-7516/2014/03/042",
    journal = "JCAP",
    volume = "03",
    pages = "042",
    year = "2014"
}

@article{Caprini:2006jb,
    author = "Caprini, Chiara and Durrer, Ruth",
    title = "{Gravitational waves from stochastic relativistic sources: Primordial turbulence and magnetic fields}",
    eprint = "astro-ph/0603476",
    archivePrefix = "arXiv",
    doi = "10.1103/PhysRevD.74.063521",
    journal = "Phys. Rev. D",
    volume = "74",
    pages = "063521",
    year = "2006"
}

@article{Fenu:2009qf,
    author = "Fenu, Elisa and Figueroa, Daniel G. and Durrer, Ruth and Garcia-Bellido, Juan",
    title = "{Gravitational waves from self-ordering scalar fields}",
    eprint = "0908.0425",
    archivePrefix = "arXiv",
    primaryClass = "astro-ph.CO",
    reportNumber = "IFT-UAM-CSIC-09-34, CERN-PH-TH-2009-145",
    doi = "10.1088/1475-7516/2009/10/005",
    journal = "JCAP",
    volume = "10",
    pages = "005",
    year = "2009"
}

@article{Moore:2014lga,
    author = "Moore, C. J. and Cole, R. H. and Berry, C. P. L.",
    title = "{Gravitational-wave sensitivity curves}",
    eprint = "1408.0740",
    archivePrefix = "arXiv",
    primaryClass = "gr-qc",
    reportNumber = "LIGO-P1400129",
    doi = "10.1088/0264-9381/32/1/015014",
    journal = "Class. Quant. Grav.",
    volume = "32",
    number = "1",
    pages = "015014",
    year = "2015"
}

@article{Turner:1993vb,
    author = "Turner, Michael S. and White, Martin J. and Lidsey, James E.",
    title = "{Tensor perturbations in inflationary models as a probe of cosmology}",
    eprint = "astro-ph/9306029",
    archivePrefix = "arXiv",
    reportNumber = "FERMILAB-PUB-93-069-A, CFPA-TH-93-19, CFPA-93-19",
    doi = "10.1103/PhysRevD.48.4613",
    journal = "Phys. Rev. D",
    volume = "48",
    pages = "4613--4622",
    year = "1993"
}

@article{Maggiore:1999vm,
    author = "Maggiore, Michele",
    title = "{Gravitational wave experiments and early universe cosmology}",
    eprint = "gr-qc/9909001",
    archivePrefix = "arXiv",
    reportNumber = "IFUP-TH-20-99",
    doi = "10.1016/S0370-1573(99)00102-7",
    journal = "Phys. Rept.",
    volume = "331",
    pages = "283--367",
    year = "2000"
}

@misc{github,
    howpublished = {\url{https://github.com/classULDM/class.VFDM}}
}

@article{BICEP:2021xfz,
    author = "Ade, P. A. R. and others",
    collaboration = "BICEP, Keck",
    title = "{Improved Constraints on Primordial Gravitational Waves using Planck, WMAP, and BICEP/Keck Observations through the 2018 Observing Season}",
    eprint = "2110.00483",
    archivePrefix = "arXiv",
    primaryClass = "astro-ph.CO",
    doi = "10.1103/PhysRevLett.127.151301",
    journal = "Phys. Rev. Lett.",
    volume = "127",
    number = "15",
    pages = "151301",
    year = "2021"
}

@article{Miron-Granese:2020hyq,
    author = "Mir{\'o}n-Granese, Nahuel",
    title = "{Relativistic viscous effects on the primordial gravitational waves spectrum}",
    eprint = "2012.11422",
    archivePrefix = "arXiv",
    primaryClass = "astro-ph.CO",
    doi = "10.1088/1475-7516/2021/06/008",
    journal = "JCAP",
    volume = "06",
    pages = "008",
    year = "2021"
}

@article{Clarke:2020bil,
    author = "Clarke, Thomas J. and Copeland, Edmund J. and Moss, Adam",
    title = "{Constraints on primordial gravitational waves from the Cosmic Microwave Background}",
    eprint = "2004.11396",
    archivePrefix = "arXiv",
    primaryClass = "astro-ph.CO",
    doi = "10.1088/1475-7516/2020/10/002",
    journal = "JCAP",
    volume = "10",
    pages = "002",
    year = "2020"
}

\end{document}